\documentclass[11pt,a4paper]{article}

\usepackage{jheppub}

\usepackage{amsmath,amssymb}
\usepackage{enumerate}
\usepackage{amsfonts}

\usepackage{subfigure}

\usepackage{allpurpose}
\usepackage{tikzrelate}

\usepackage{tikz}

\usepackage[marginal,perpage,symbol]{footmisc}
\usepackage{slashed} %

\makeatletter
\gdef\@fpheader{}
\makeatother

\usepackage[nottoc,notlof,notlot]{tocbibind}

\usepackage{indentfirst}

\begin{document}

\title{Late-time cosmic acceleration from quantum gravity}
\date{\today}

\author[*,a]{Daniele Oriti\note{Corresponding author.}}
\affiliation[a]{Departamento de F\'{i}sica Te\'{o}rica and IPARCOS, Facultad de Ciencias F\'{i}sicas, Universidad Complutense de Madrid, Plaza de las Ciencias 1, 28040 Madrid, Spain, EU}  
\emailAdd{doriti@ucm.es}

\author[b]{and Xiankai Pang}
\affiliation[b]{School of Physics and Astronomy, China West Normal University, Shida Road 1, 637009, China}
\emailAdd{xkpang@cwnu.edu.cn}

\abstract{We deepen the analysis of the cosmological acceleration produced by quantum gravity dynamics in the formalism of group field theory condensate cosmology, treated at the coarse-grained level via a phenomenological model, in the language of hydrodynamics on minisuperspace. Specifically, we conduct a detailed analysis of the late-time evolution, which shows a phantom-like phase followed by an asymptotic De Sitter expansion. We argue that the model indicates a recent occurrence of the phantom crossing and we extract a more precise expression for the effective cosmological constant, linking its value to other parameters in the model and to the scale of the quantum bounce in the early universe evolution. Additionally, we show how the phantom phase produced by our quantum gravity dynamics increases the inferred value of the current Hubble parameter based on observed data, indicating a possible quantum gravity mechanism for alleviating the Hubble tension. Our results represent a concrete example of how quantum gravity can provide an explanation for large-scale cosmological puzzles, in an emergent spacetime scenario.}


\maketitle

\section{Introduction}

Any complete quantum gravity theory must reproduce conventional spacetime descriptions governed by General Relativity (GR)~\cite{Novikov:2016fzd} and effective quantum field theory (QFT). This requirement proves more straightforward for approaches building on established field-theoretic structures (e.g., asymptotic safety~\cite{Reuter:2019byg} or string theory~\cite{Blau:2009fzi}) compared to frameworks postulating fundamentally non-spatiotemporal entities, such as tensorial group field theories (TGFTs) \cite{Krajewski:2012aw,Carrozza:2016vsq,Oriti:2014uga,Rivasseau:2012yp, Rivasseau:2016zco, Rivasseau:2016wvy, Carrozza:2024gnh}, and a specialized class of them, often called group field theories (GFTs)~\cite{Oriti:2009wn}. The latter face the significant challenge of spacetime emergence \cite{Oriti:2018dsg}, where difficulty increases with the conceptual distance between fundamental constituents and classical spacetime notions.

One way to ease the difficulty of recovering classical spacetime is to focus on spacetimes with certain symmetries, such as spherically symmetric~\cite{Oriti:2015rwa,Oriti:2018qty,Oriti:2023yjj} and homogeneous~\cite{Gielen:2013naa} spacetimes. The latter leads to 
GFT condensate cosmology \cite{Gielen:2013naa,Oriti:2016qtz,Gielen:2016dss, Oriti:2016acw,Pithis:2019tvp,Marchetti:2020qsq,Marchetti:2020umh,Oriti:2024qav}, which seeks cosmological predictions from GFTs by studying hydrodynamic approximations of quantum dynamics, particularly focusing on condensate states. This approach models the universe as a quantum fluid of GFT quanta. Recent work demonstrates that mean-field approximations can yield physically significant results, including derivation of hydrodynamics on minisuperspace \cite{Oriti:2024qav} -- a framework connecting quantum gravity effects to cosmology. Our previous analysis \cite{Oriti:2021rvm} showed how late-time cosmic acceleration emerges naturally in this scheme. The present work extends these results through detailed examination of the acceleration phase, especially the modifications to the $\Lambda$CDM model.

As a standard model in cosmology, the $\Lambda$CDM model is very successful in describing the evolution of our universe using only a few parameters~\cite{Aghanim:2018eyx}, and the cosmological constant can indeed be seen as just another constant of nature as far as classical GR is concerned, contributing to the vacuum energy of the universe. However, the magnitude of the observed cosmological constant is much less than what one would expect, either compared to the vacuum energy from particle physics~\cite{Martin:2012bt}, or to the value that would be obtained from the renormalization flow of gravity~\cite{Weinberg:1988cp}. This is one traditional way of seeing the cosmological constant problem~\cite{Weinberg:1988cp}. 
There are many ideas to address the problem formulated in this way. For example, one may object that the particle vacuum energy shouldn't be the source of $\Lambda$, as what can be observed is differences between energies and not the absolute energy itself~\cite{Bianchi:2010uw}. Or in unimodular gravity (classically equivalent to GR), we can take $\Lambda$ as an integral constant that is not subject to quantum fluctuations~\cite{Eichhorn:2013xr, Percacci:2017fsy}. More generally, one could be wary of importing straightforward QFT reasoning in the gravitational sector, since we lack a coherent framework for quantum fields coupled to quantum gravity, or argue that only a proper non-perturbative treatment of gravity in the quantum domain could clarify fully the physics of the cosmological constant.

However, despite its successes, the $\Lambda$CDM model faces mounting challenges from recent observational data. Supernova studies, for instance, suggest the possibility of a current phantom phase~\cite{Shafer:2013pxa, Zhao:2017cud, Wang:2018fng}, characterized by an equation of state (EoS) parameter \( w < -1 \), a scenario inconsistent with the $\Lambda$CDM framework. Further tension arises from the Hubble constant (\( H_0 \)): values inferred from the CMB~\cite{Aghanim:2018eyx} are systematically lower than those derived from local distance-ladder measurements~\cite{Riess:2019cxk, Perivolaropoulos:2021jda}. This \( H_0 \) discrepancy has been interpreted as potential evidence for a late-time phantom phase in cosmic expansion~\cite{DiValentino:2020naf, Yang:2021hxg, Heisenberg:2022gqk,Lee:2022cyh}. 
 Although the recent DESI results indicate that the current value of the EoS parameter \( w \) should be greater than $-1$~\cite{DESI:2024mwx,DESI:2025zgx}, the presence of a phantom phase in the evolution history is also necessary~\cite{DESI:2025zgx}, which suggests that the dark energy might be dynamical.
Collectively, these findings underscore the need to revise the standard cosmological expansion history, particularly its late-time evolution~\cite{Vagnozzi:2019ezj, Vagnozzi:2023nrq}, going beyond a constant vacuum energy component, i.e. a cosmological constant. They represent an important motivation for our work.

In previous work, we have shown that a late-time accelerating expansion can emerge rather naturally from quantum gravity theory, in GFT condensate cosmology ~\cite{Oriti:2021rvm}. In fact, one can obtain a phantom-like phase without introducing any phantom field, but from the purely quantum gravity effects, in an emergent spacetime scenario. 
 Even though recent DESI results favor \( w > -1 \) at low redshifts~\cite{DESI:2024mwx,DESI:2025zgx}, the possibility of systematic discrepancies between observational datasets cannot yet be ruled out~\cite{Colgain:2024mtg}. Therefore, it remains legitimate to consider models in which \( w < -1 \) today, as is the case in our GFT framework.

In this contribution, we take a further step toward bridging our previous results with cosmological observations, analysing in more detail the cosmological dynamics extracted from the quantum gravity model. 
In particular,  among other findings, a) we extract the precise expression for the asymptotic cosmological constant as a function of the parameters of the model and of the condensate state; b) we show that the phantom phase will not last long (in cosmological scales) before the universe enters a de sitter regime (hence, if we are experiencing a phantom phase now, the phantom crossing must have happened recently); c) we extract some observational consequences of the phantom phase, and using methods from~\cite{Heisenberg:2022gqk}, we show that its occurrence increases $H_0$, the current value of Hubble parameter, inferred from cosmological evolutions. 

As in our previous work, we also focus on the two-modes\footnote{The precise meaning of 'modes' will become clear when we introduce the GFT formalism in section \ref{sec:gftcosreview}.} correction to the single-mode condensate, where the additional mode will introduce a phantom phase (\acrshort{eos} $w<-1$) to the evolution. In principle, one can also include more than two modes of the GFT condensate in the dynamics, but the asymptotic behaviour (see equation \eqref{eq:largerhosol}) of the interacting GFT condensate shows that other modes will not alter the phantom behaviour, just as the second mode will not modify the asymptotic de Sitter phase (where $w=-1$) determined by the single mode model. This is because the condensate dynamics leads to the progressive late-time dominance of the lowest condensate modes.

{It's also worth emphasising that we are still a step away from connecting our results to the actual cosmological observations. The main limitations are, first, that currently we are not able to incorporate matter fields such as radiation and non-relativistic matter into the GFT formalism, without which we cannot model the cosmological evolution in full detail with realistic matter components, and thus compare with the $\Lambda$CDM model; second, for similar reasons, we are not yet able to provide precise estimates of the redshift at which cosmological events take place. None of these two obstacles is a matter of principle, but rather a temporary technical difficulty, thus we think that the gap can be closed in the near future. Nevertheless, the late-time evolution is dominated by the cosmological constant rather than the non-relativistic matter, and hence in late time we can focus on GFT and ignore other kinds of matter fields. Our result shows that in the GFT formalism, the cosmological evolution contains a phantom phase which increases $H_0$.}

The paper is organized as follows. In section \ref{sec:gftcosreview}, we give a brief introduction of GFT and its cosmological sector, described by hydrodynamics on minisuperspace. Section \ref{sec:effwsingle} summaries our previous results on the behaviour of effective equation of state (EoS). Then in section \ref{sec:effwbehaviour}, we discuss how to obtain the minimal value of EoS, i.e. the deep phantom regime and its location. The last part of section \ref{sec:effwbehaviour} shows the relation between the effective cosmological constant and parameters of the quantum gravity GFT. Section \ref{sec:deviationsingle} proposes a definition of redshift $z$ in the GFT formalism, which allows us to analyse more quantitatively the expansion history of the universe and the effects on the current value $H_0$ of the Hubble parameter. Finally, in section \ref{sec:summary}, we summarize our result and point out several possible directions for future investigations.

Throughout the manuscript, natural units ($c=\hbar=1$) are used unless specified.

\section{GFT condensate cosmology} \label{sec:gftcosreview}
In this section, we provide a brief overview of the basics of the TGFT formalism, with a focus on quantum geometric models (i.e., GFTs) for $4$d quantum gravity, specifically the elements upon which the extraction of cosmological dynamics is based. We aim to include only the key ingredients necessary for the immediate context of this work. For a more detailed introduction to TGFTs, we refer the reader to existing reviews \cite{Freidel:2005qe, Krajewski:2012aw,Oriti:2014uga}. For the foundational aspects of GFT cosmology, we recommend the original works \cite{Gielen:2013kla,Gielen:2013naa,Oriti:2016qtz,Oriti:2016ueo}, along with the reviews \cite{Gielen:2016dss,Oriti:2016acw,Pithis:2019tvp}. Additionally, works such as \cite{Marchetti:2020umh,Marchetti:2020qsq} explore the use of coherent peaked states for studying relational observables and their dynamics, which are relevant to the current research.

\subsection{Group field theory formalism}

\begin{figure}[htp]
  \centering
  \subfigure[$4$-valent vertex]{\label{fig:vertex}\begin{tikzpicture}[scale=0.7,transform shape,every node/.style={font={\large}}]

  \draw (0.2,-0.3) node[circle, fill, inner sep=2pt,minimum size=2pt] (orig) {};
  \draw[-{Latex[length=3.5mm, width=1.4mm]}] (orig) -- (0.7,2.6) node[right] (j4) {$j_4$};
  \draw[-{Latex[length=3.5mm, width=1.4mm]}] (orig) -- (2.2,-1.9) node[below] (j3) {$j_3$};
  \draw[-{Latex[length=3.5mm, width=1.4mm]}] (orig) -- (-2.5,-0.9) node[below] (j2) {$j_2$};
  \draw[-{Latex[length=3.5mm, width=1.4mm]}] (orig) -- (-3.5,0.3) node[above] (j1) {$j_1$};
  \draw (0,-4) node[] {};
  \draw (0.2,-0.4) node[below] {$\displaystyle \iota$};

\end{tikzpicture}}
  \subfigure[Dual tetrahedron]{\label{fig:tetrahedra}\begin{tikzpicture}[scale=0.7,transform shape,every node/.style={font={\large}}]

  \draw (0.2,-0.3) node[circle, fill, inner sep=2pt,minimum size=2pt] (orig) {};

  \draw (-3.464,2) node[inner sep=0pt,outer sep=0pt] (o1) {};
  \draw (3.464,2) node[inner sep=0pt,outer sep=0pt] (o2) {};
  \draw (0,-4) node[inner sep=0pt,outer sep=0pt] (o3) {};
  \draw (1,1) node[inner sep=0pt,outer sep=0pt] (o4) {};

  \draw (o1) -- ++ (o2) -- ++ (o3) -- ++ (o1) -- ++ (o4) -- ++ (o2);
  \draw (o4) -- ++ (o3);

  \draw[dashed] (orig) -- (0.5,1.4) node[inner sep=0pt] (j4h) {};
  \draw[-{Latex[length=3.5mm, width=1.4mm]}] (j4h) -- (0.7,2.6) node[right] (j4) {$j_4$};
 \draw[dashed] (orig) -- (1.2,-1.1) node[inner sep=0pt] (j3h) {};
  \draw[-{Latex[length=3.5mm, width=1.4mm]}] (j3h) -- (2.2,-1.9) node[below] (j3) {$j_3$};
 \draw[dashed] (orig) -- (-0.9,-0.5) node[inner sep=0pt] (j2h) {};
  \draw[-{Latex[length=3.5mm, width=1.4mm]}] (j2h) -- (-2.5,-0.9) node[below] (j2) {$j_2$};
  \draw[dashed] (orig) -- (-2.3,0.1) node[inner sep=0pt] (j1h) {};
  \draw[-{Latex[length=3.5mm, width=1.4mm]}] (j1h) -- (-3.5,0.3) node[above] (j1) {$j_1$};
  \draw (0.2,-0.4) node[below] {$\displaystyle \iota$};
\end{tikzpicture}}
  \caption[$4$-valent vertex and its dual tetrahedron]{$4$-valent vertex of a spin network and its dual tetrahedron. Each link $i$ incident to the vertex is associated with a spin $j_i$ of the $SU(2)$ group, and the vertex itself is labelled by the intertwiner $\iota$; they determine the areas of the corresponding triangular faces and the volume of the tetrahedron, respectively.}
	  \label{fig:vertextetrahedra}
\end{figure}
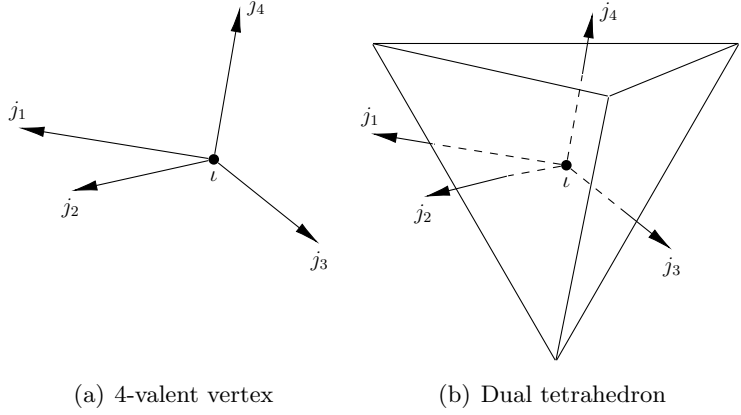

GFTs are quantum field theories (QFTs) defined on multiple copies of a Lie group $G$, replacing the spacetime manifold that serves as the base in conventional QFTs. These theories encode, through their fields and conjugate variables, the quantum geometric data required to reconstruct spacetime geometry and its physical degrees of freedom. In $4$d quantum gravity models, for instance, the fundamental field is a tensorial complex-valued function $\varphi:~G^{\times 4} \to \mathbb{C},~\varphi(g_v) = \varphi(g_1, \cdots, g_4)$, where the rank of the tensor (here $4$) corresponds to the spacetime dimension being modeled \cite{Krajewski:2012aw}. Crucially, GFTs are formulated as field theories \emph{of} spacetime—describing the kinematics and dynamics of its fundamental constituents—rather than QFTs \emph{on} a pre-defined spacetime background. 
The elementary quanta of a GFT are combinatorial 3-simplices (tetrahedra) labeled by group-theoretic data (see figure \ref{fig:vertextetrahedra}), which encode their intrinsic quantum geometry. These tetrahedra are typically interpreted as spacelike building blocks of quantum geometry. Quantum states and boundary conditions in these models are then represented by configurations of such quanta, analogous to Fock states in many-body quantum systems.

In existing Lorentzian-signature quantum gravity models, { including the much studied EPRL model as well as the BC model ~\cite{Oriti:2016qtz,Jercher:2021bie,Marchetti:2022igl}}\footnote{We refer here to their GFT formulation, which provide a completion of the spin foam amplitudes for individual complexes, since this is the language used in the following analysis.}, the group 
$G$ is typically chosen as the double cover of the Lorentz group, i.e., $G=SL(2, \mathbb{C})$, or its rotation subgroup $SU(2)$, which arises in symmetry-reduced or Euclideanized settings. Dynamical constraints in these models ensure that the group-theoretic data admit a geometric interpretation, often aligning with discrete analogues of Einstein’s equations or spin foam dynamics. This allows mappings between formulations based on $SL(2,\mathbb{C})$ and $SU(2)$,  despite their differing geometric roles \cite{Engle:2007wy, Engle:2007uq, Dupuis:2010jn, Finocchiaro:2020xwr}. For greater simplicity and to align with widely studied models, we adopt $G=SU(2)$ in the following discussion. Further details on geometric constraints and group-theoretic interpretations can be found in the cited literature.

\paragraph{Second quantization.} The field $\varphi(g_v)$ must satisfy geometric constraints that ensure its consistency with the quantum geometry of spacetime. Specifically, the field is required to be right-invariant, meaning that it satisfies the condition \cite{Oriti:2016qtz}
\iea{
\varphi(g_v h) = \varphi(g_1 h, g_2 h, g_3 h, g_4 h) = \varphi(g_v), \quad \forall h \in G,
}
which implies that $\varphi(g_v) \in L^2(G^{\times 4}/G)$. This restriction is essential for the correct representation of spacetime in the GFT formalism.
Then we use a complete and orthonormal basis for the Hilbert space $L^2(G^{\times 4}/G)$. This basis is constructed from $SU(2)$ Wigner representation functions that are contracted by group intertwiners, resulting in what are known as spin network vertex functions. 
At the second-quantised\footnote{This second-quantized formulation is not directly obtained by quantizing the classical GFT action using standard canonical methods, due to the absence of an external time parameter, which is typically needed for such methods. However, this timeless formalism can be understood in terms of standard canonical quantization techniques from a "frozen" perspective \cite{Gielen:2021dlk}, where the GFT model is viewed as a constrained system. Alternative "deparametrized" canonical formulations of the same GFTs (after incorporating additional "matter" degrees of freedom, as discussed in later sections) also exist \cite{Wilson-Ewing:2018mrp, Gielen:2019kae}. For an insightful recent review of this alternative construction of a Hilbert space structure for GFTs, see \cite{Gielen:2024sxs}.} level, the field $\varphi(g_v)$ can be prompted to operators as following~\cite{Oriti:2013aqa}
\iea{
  \hat{\varphi}(g_v) = \sum_{\vec{x}} \hat{c}_{\vec{x}} \kappa_{\vec{x}}(g_v), \quad \hat{\varphi}^\dag(g_v) = \sum_{\vec{x}} \hat{c}_{\vec{x}}^\dag \bar{\kappa}_{\vec{x}}(g_v),
}
where $\kappa_{\vec{x}}(g_v)$ are basis functions~\cite{Oriti:2013aqa} whose exact form will not concern us in this work. The annihilation operator $\hat{c}_{\vec{x}}$ and creation operator $\hat{c}_{\vec{x}}^\dag$ act to create and annihilate spin network nodes (or equivalently, tetrahedra) labeled by $\vec{x} = (\vec{j}, \vec{m}, \iota)$. These operators satisfy the following commutation relations:
\iea{
  \left[\hat{c}_{\vec{x}}, \hat{c}^\dag_{\vec{x}'}\right] = \delta_{\vec{x}, \vec{x}'}, \quad \left[\hat{c}_{\vec{x}}, \hat{c}_{\vec{x}'}\right] = \left[\hat{c}^\dag_{\vec{x}}, \hat{c}^\dag_{\vec{x}'}\right] = 0.
}

The Fock vacuum state $\ket{0}$, which represents a state with no spacetime structure (neither geometrical nor topological), is defined by the condition $\hat{c}_{\vec{x}} \ket{0} = 0$ for all $\vec{x}$. By acting repeatedly with the creation operator $\hat{c}^\dag_{\vec{x}}$ on the vacuum state, we can construct the many-body states %
and lead to the extended topological structures,%
which captures the pattern of entanglement between the fundamental degrees of freedom \cite{Colafranceschi:2020ern}.

Next, we introduce the volume operator, which in its first quantized form is diagonal in the spin network basis, with matrix elements that depend on the intertwiner label $\iota$. The second quantized version of the volume operator is obtained by convolving the matrix elements of the first quantized operator with field operators, as is customary in quantum many-body systems. We can express the volume operator as:
\iea{
  \hat{V} = \sum_{\vec{x}, \vec{x}'} V({\vec{j}};\iota, \iota') \delta_{\vec{x} - \{\iota\}, \vec{x}' - \{\iota'\}} \hat{c}_{\vec{x}}^\dag \hat{c}_{\vec{x}'}.
}

This construction follows the same pattern as the other quantum geometric and matter operators in the theory.

\paragraph{Coupling to a scalar field.} 
In quantum gravity models, incorporating matter degrees of freedom is essential for ensuring physical viability. In diffeomorphism-invariant contexts, a relational approach to defining time evolution is often employed. This method utilizes internal dynamical degrees of freedom as a clock, with the evolution of other degrees of freedom defined relative to it. In many applications, particularly in GFT cosmology, a massless, non-interacting scalar field serves this clock role.

To introduce scalar matter into the formalism, we consider a single scalar field. These scalar field degrees of freedom are integrated with the quantum geometric ones in the fundamental definition of the GFT model. The first step involves extending the GFT field to a map $\varphi: G^{\times 4} \times \mathcal{R} \to \mathcal{C}$, and subsequently, the GFT action is modified to include appropriate couplings of the new degrees of freedom. The primary guideline for constructing such extended dynamics mirrors that of pure geometry models: the GFT model is defined so that its perturbative expansion yields a sum over simplicial complexes, each weighted by a discrete path integral for gravity, now coupled to a massless, non-interacting scalar field \cite{Gielen:2013naa,Oriti:2016qtz,Li:2017uao}.

It's worth emphasizing that, although the interpretation of these new degrees of freedom—similar to the quantum geometric ones—is shaped by their behaviour at the discrete level, such as through GFT quanta and Feynman amplitudes, their true physical significance and characteristics should be understood in terms of their effective roles in a continuum framework. The goal of the GFT cosmology program is to derive these effective descriptions and gain a deeper understanding of the emergent physics within these quantum gravity models.

The dependence on the relational time $\phi$ will be kept as well after the quantization. The commutation relation between annihilation and creation operators then read:
\[
\left[\hat{c}_{\vec{x}}(\phi), \hat{c}^\dag_{\vec{x}'}(\phi')\right] = \delta_{\vec{x}, \vec{x}'} \delta(\phi' - \phi), \quad \left[\hat{c}_{\vec{x}}(\phi), \hat{c}_{\vec{x}'}(\phi')\right] = \left[\hat{c}^\dag_{\vec{x}}(\phi), \hat{c}^\dag_{\vec{x}'}(\phi')\right] = 0.
\]

Similarly, the definitions of other observables will also account for their dependence on the scalar field degrees of freedom. For instance, the volume operator, which sums the contributions from each GFT quantum, takes the following form:
\begin{equation}
\hat{V} = \int \dd\phi \, \hat{V}(\phi) = \int \dd\phi \sum_{\vec{x}, \vec{x}'} V({\vec{j}};\iota, \iota') \delta_{\vec{x} - \{\iota\}, \vec{x}' - \{\iota'\}} \hat{c}_{\vec{x}}^\dag(\phi) \hat{c}_{\vec{x}'}(\phi).
\end{equation}

The relational approach suggests defining a relational observable that corresponds to the volume of the universe at a \emph{specific clock time}, with the scalar field acting as the clock. One possible initial definition is given by the quantity $\hat{V}(\phi)$, as shown in the expression above. This definition has been used in several parts of the GFT cosmology literature. More recently, an effective relational strategy has been introduced, where relational observables are identified as the expectation values of generic GFT operators within appropriately chosen "clock-peaked" states \cite{Marchetti:2020umh}. We will present this effective strategy in more detail after discussing the dynamical aspects of the theory. For a comprehensive construction of relational observables within the GFT framework, and to understand the connections between different definitions, we refer to \cite{Marchetti:2024nnk}.

{Before we proceed, let us briefly comment on the role of the massless scalar field in the cosmological evolution. In our model, the evolution is determined by both the matter content and quantum gravity effects~\cite{Oriti:2016qtz,Oriti:2021rvm}. Following an initial bounce dominated by quantum effects, the universe enters a FLRW phase, where the dynamics are governed by the massless scalar field, which has an \acrshort{eos} $w = 1$. At later times, quantum gravity effects become dominant again, giving rise to a phantom phase and an asymptotic de Sitter phase characterized by $w = -1$~\cite{Oriti:2021rvm}.

Changing the relational clock---that is, replacing the massless scalar field with another matter field---modifies the matter content of the universe and alters the intermediate evolution. Nevertheless, the quantum effects are expected to dominate at late times, as long as the matter field has an equation of state parameter $w > 0$. The inclusion of matter fields in the \acrshort{gft} formalism is generally a challenging task. Therefore, in this work, we restrict ourselves to using a massless scalar field as a relational clock.
}

\paragraph{Dynamics.} Classically, the dynamics of a given GFT model are encoded in the action
\iea{
  S(\bar{\varphi},\varphi)&=& \int\dd g_{v_1}\dd g_{v_2}\bar{\varphi}(g_{v_1})\varphi(g_{v_2})K(g_{v_1},g_{v_2})\nonumber \\
  &&-\sum_{n,m}^\infty\lambda_{n+m}\int\left[(\dd g_v)^{m}(\dd h_v)^{n}\prod_{i=1}^m\bar{\varphi}(g_{v_i})\prod_{j=1}^n\varphi(h_{v_j})V_{n+m}(g_{v},h_{v})\right],
}
where $K(g_{v_1}, g_{v_2})$ and $V_{n+m}(g_v,h_v) = V_{n+m}(g_{v_1}, \dots, g_{v_m}, h_{v_1}, \dots, h_{v_n})$ are the kinetic and interaction kernels, respectively. We use a notation inspired by quantum many-body physics, which reflects the possibility of various interactions involving different numbers of 'spacetime atoms'. For simplicity, we focus on the case where only pure quantum geometric data are considered. In this context, the interaction kernels are generally non-local with respect to the quantum geometric data, meaning that the field arguments are not simply identified in these kernels. However, for the scalar field degrees of freedom are introduced, the interaction kernels are local as usual in QFTs.

To get the quantum dynamics from the classical ones, one can consider the partition function:
\ieas{
  Z=\int\mathcal{D}\varphi\mathcal{D}\bar{\varphi}\ee^{-S(\bar{\varphi},\varphi)},
}
from which the quantum equation of motion, i.e., the Schwinger-Dyson equations can be derived by taking variations respect to the expectation value of any operator $O(\bar{\varphi},\varphi)$ \cite{Gielen:2013naa, Gielen:2016dss}:
\iea{
  0=\int\mathcal{D}\varphi\mathcal{D}\bar{\varphi}\frac{\delta}{\delta\bar{\varphi}}\left(O(\bar{\varphi},\varphi)\ee^{-S(\bar{\varphi},\varphi}\right)=\left\langle\frac{\delta O(\bar{\varphi},\varphi)}{\delta \bar{\varphi}}-O(\bar{\varphi},\varphi)\frac{\delta S(\bar{\varphi},\varphi)}{\delta \bar{\varphi}}\right\rangle,
}
where the vacuum expectation value $\langle \cdots \rangle$ is defined as:
\ieas{
  \langle O(\bar{\varphi},\varphi)\rangle=\int\mathcal{D}\varphi\mathcal{D}\bar{\varphi}O(\bar{\varphi},\varphi)\ee^{-S(\bar{\varphi},\varphi)}.
}

In the case of small quantum fluctuations, a mean-field approximation is expected to be valid. At leading order, we only need to consider the simplest equation in the series of Schwinger-Dyson equations:
\ieas{
  \left\langle\sigma\left|\frac{\delta \hat{S}(\hat{\varphi}^\dag,\hat{\varphi})}{\delta\hat{\varphi}^\dag}\right|\sigma\right\rangle=0, 
}
where $\ket{\sigma}$ is expected to be an arbitrary state. In particular, the equations valid for eigenstates of the field operator, i.e., we can take $\ket{\sigma}$ such that $\hat{\varphi}(g_v, \phi) \ket{\sigma} = \sigma(g_v, \phi) \ket{\sigma}$ for some eigenfunction $\sigma(g_v, \phi)$. And then the dynamics can be expressed as an equation of motion for $\sigma(g_v, \phi)$, which is derived from the effective action:
\begin{equation}
  S(\bar{\sigma}, \sigma) = \langle \sigma | S(\hat{\varphi}^\dag, \hat{\varphi}) | \sigma \rangle. \label{MeanFieldDyn}
\end{equation}

This effective dynamics, in terms of the GFT mean-field, encoded in the action (\eqref{MeanFieldDyn}), takes the form of hydrodynamics on minisuperspace \cite{Oriti:2024qav}, due to the isomorphism between the domain of the GFT field (in quantum geometric models) and minisuperspace (or the space of geometries + matter fields at a point in continuum physics) \cite{Gielen:2014ila, Jercher:2021bie}.

{In practice, going beyond the hydrodynamic approximation also requires accounting for quantum fluctuations. The effective action~\eqref{MeanFieldDyn} implicitly assumes that
\[
\braket{\sigma|\hat{c}^{\dag}_{\vec{x}}(\phi)\hat{c}_{\vec{x}}(\phi)|\sigma} = \braket{\sigma|\hat{c}^{\dag}_{\vec{x}}(\phi)|\sigma} \braket{\sigma|\hat{c}_{\vec{x}}(\phi)|\sigma},
\]
which holds only for coherent states. For the true ground state $\ket{\Omega}$, however, one has
\begin{IEEEeqnarray}{c}
    S = \braket{\Omega|S(\hat{\varphi}^\dag, \hat{\varphi}) |\Omega} = S(\bar{\sigma}, \sigma) + \chi^2,
\end{IEEEeqnarray}
where $\chi$ represents the quantum corrections arising from fluctuations. These fluctuations, however, are expected to be suppressed in the presence of a large number of \acrshort{gft} quanta, which are required to recover the continuum limit. Moreover, for the ground state $\ket{\Omega}$ to remain stable, we also expect $\chi^2$ to be suppressed over time. Therefore, at leading order, the hydrodynamic approximation remains valid for the purposes of this work. A detailed analysis of quantum fluctuations is certainly important, but lies beyond the scope of the present manuscript and will be addressed in future works.
}

In the following, we will explore how this approximation plays out for a specific class of condensate wavefunctions, leading to an effective definition of relational observables and the emergence of cosmological dynamics.

\subsection{GFT condensate cosmology}

\paragraph{Coherent peaked states.} %
In the framework of GFT cosmology, the evolution of the universe is captured by the transformation of a spatial slice of spacetime, where this transformation occurs with respect to relational time $\phi$. To incorporate the dependence of observables on the clock variable, we focus on states that are sharply peaked around a specific relational time $\phi_0$ \cite{Marchetti:2020umh}. Furthermore, it's also necessary for these states to support the appropriate observables such that the continuum spacetime can emerge under suitable limits  \cite{Oriti:2016qtz}. These requirements naturally lead to the adoption of coherent peaked states (CPS), which provide an effective framework for describing the dynamics at the quantum geometric level:
\iea{
  \ket{\sigma_\varepsilon;\phi_0,\pi_0}=\mathcal{N}(\sigma)\exp\left(\int(\dd g)^4\dd\phi \sigma_\varepsilon(g_v,\phi;\phi_0,\pi_0)\hat{\varphi}^{\dag}(g_v,\phi)\right)\ket{0}, \label{eq:annidef}
}
with $\mathcal{N}(\sigma)$ is a constant responsible to achieve the correct normalization and $\ket{0}$ is the vacuum state. The condensate wavefunction $\sigma_\varepsilon(g_v,\phi;\phi_0,\pi_0)$ is peaked on $\phi=\phi_0$ and can be written as \cite{Marchetti:2020qsq}
\iea{
  \sigma_\varepsilon(g_v,\phi;\phi_0,\pi_0)=\eta_\varepsilon(\phi-\phi_0,\pi_0)\tilde{\sigma}(g_v,\phi),
}
where $\eta_\varepsilon(\phi-\phi_0,\pi_0)$ represents a \emph{peaking function}, typically chosen as a Gaussian (see equation (52) in \cite{Marchetti:2020umh}), centered around $\phi_0$ with a characteristic width controlled by $\varepsilon$. The parameter $\pi_0$ further governs the fluctuations of the operator corresponding to the conjugate momentum of the scalar field $\phi$. The \emph{reduced condensate function} $\tilde{\sigma}(g_v,\phi)$, which serves as the primary dynamical variable in the hydrodynamic approximation, does not alter the peaking behavior of the function $\sigma_\varepsilon(g_v,\phi;\phi_0,\pi_0)$, which is dictated by the peaking function $\eta_\varepsilon(\phi-\phi_0,\pi_0)$. Furthermore, it remains true that the condensate state defined in \eqref{eq:annidef} continues to be an eigenstate of the GFT field operator such that 
\iea{
  \hat{\varphi}(g_v,\phi)\ket{\sigma_\varepsilon;\phi_0,\pi_0}=\sigma_\varepsilon(g_v,\phi;\phi_0,\pi_0)\ket{\sigma_\varepsilon;\phi_0,\pi_0}.
}

Furthermore, motivated by geometric considerations, we see that the condensation wave condition should be invariant under both right and left diagonal group actions~\cite{Marchetti:2020umh, Gielen:2013naa, Oriti:2016qtz,Gielen:2014ila}, i.e., we need impose the following condition on $\sigma(g_v,\phi)$ as well
\iea{
  \tilde{\sigma}(h g_v k,\phi)=\tilde{\sigma}(g_v,\phi),~\forall ~h,~k\in SU(2).
}

\paragraph{Imposing isotropy.}
Recall that our goal is to derive the cosmological dynamics of homogeneous and isotropic universes from the hydrodynamics of the GFT condensate, with the evolution of the universe's volume serving as the primary observable. To achieve this, we introduce an additional constraint on the condensate wavefunction, i.e., isotropy.

From the perspective of discrete geometry, as encoded in the GFT Fock space, this implies that the wavefunction $\sigma(g_I, \phi)$ must only have support over equilateral tetrahedra~\cite{Oriti:2016qtz}. This condition enforces the requirement that the spin labels (which correspond to the areas of the boundary triangles) be equal, and the intertwiners should {be chosen such that the volume takes the largest value} allowed by this choice of spins. Furthermore, taking into account both left and right invariance, the condensate wavefunction then assumes to be~\cite{Oriti:2016qtz}
\iea{
  \tilde{\sigma}(g_I,\phi)=\sum_j\tilde{\sigma}_j(\phi)\bar{\mathcal{I}}^{j,\iota_{+}}_{\vec{m}}\mathcal{I}^{j,\iota_{+}}_{\vec{n}}d(j)^2\prod_{l=1}^4D^j_{m_ln_l}(g_l), \label{eq:sigmadecomp}
}
here $j$ is a shorthand notation the collection of spins $\vec{j}=(j_1,j_2,j_3,j_4)=(j,j,j,j)$, and similarly for $\vec{m},~\vec{n}$; 
$\mathcal{I}^{j, \iota_+}_{\vec{m}}$ represents the intertwiner labeled by $\iota$, while $d(j) = 2j + 1$ is the dimension of the spin-$j$ representation. The functions $D^{j}_{m_l n_l}(g_l)$ are the Wigner representation functions. After these considerations, the time evolution of the isotropic state is solely encoded in the function $\sigma_j(\phi)$ for each mode $j$. Additionally, recalling the definition of the annihilation operator in equation \eqref{eq:annidef}, we have:
\iea{
  \hat{c}_{\vec{x}}(\phi)\ket{\sigma_\varepsilon;\phi_0,\pi_0}=\eta_\varepsilon(\phi-\phi_0,\pi_0)\tilde{\sigma}_j(\phi)\bar{\mathcal{I}}^{j,\iota_+}_{\vec{m}}\ket{\sigma},
}
i.e., the action of $\hat{c}_{\vec{x}}$ with $\vec{x}=(\vec{j},\vec{m},\iota)$ is non-vanishing only when the $4$ spins are identical $j_1=j_2=j_3=j_4=j$, as one should expect by the construction of the isotropic states. 

\paragraph{Effective dynamics.}

Since the peaking function $\eta_\varepsilon(\phi - \phi_0, \pi_0)$ is usually chosen as a Gaussian, the time evolution of the condensate can be traced once the reduced condensate function $\tilde{\sigma}(g_v, \phi)$ is known. At the mean-field level, the evolution of $\tilde{\sigma}(g_v,\phi)$ is encoded in an effective action, which takes the following form~\cite{Marchetti:2020umh}:
\iea{
 S(\bar{\tilde{\sigma}},\tilde{\sigma})&=&\int\dd\phi_0\langle\sigma_\varepsilon;\phi_0,\pi_0|S(\hat{\varphi}^\dag,\hat{\varphi})|\sigma_\varepsilon;\phi_0,\pi_0\rangle, \nonumber \\
  &=&\int\dd\phi_0\left\{\sum_j\left[\bar{\tilde{\sigma}}_j(\phi_0)\tilde{\sigma}''(\phi_0)-2\ii \tilde{\pi}_0\bar{\tilde{\sigma}}_j(\phi_0)\tilde{\sigma}_j'(\phi_0)-\xi_j^2\bar{\tilde{\sigma}}_j(\phi_0)\tilde{\sigma}_j(\phi_0)\right]+\mathcal{V}(\bar{\tilde{\sigma}},\tilde{\sigma})\right\}, \nonumber \\
  \label{eq:condaction}
}
where $\displaystyle \tilde{\pi}_0 = \frac{\pi_0}{\varepsilon \pi_0^2 - 1}$, and $\xi_j$ is an effective parameter that encodes the details of the kinetic term in the fundamental GFT action under the isotropic restriction. The derivatives denoted by $'$ represent derivatives with respect to $\phi_0$. Finally, $\mathcal{V}(\bar{\tilde{\sigma}}, \tilde{\sigma})$ is the interaction kernel, which is also determined by the underlying GFT model. For further details, please refer to \cite{Marchetti:2020umh} and the references therein.

The interaction term in GFT models usually has complicate forms, even under the isotropic restriction, and the corresponding dynamics are challenging to manage, even at the mean-field level. For practical reasons, many studies have previously neglected the contribution from these interaction terms, assuming them to be subdominant compared to the kinetic term\footnote{This assumption is also necessary for the perturbative formulation of the GFT quantum dynamics, where the connection to spin foam models and the lattice gravity path integral becomes relevant.}. In contrast, the focus of this work is precisely on understanding how these interaction terms influence the effective cosmological dynamics, particularly at late times.

To this end, we take a phenomenological approach, where we model the interactions using a general and simplified form. This approach, which has been utilized in previous studies \cite{deCesare:2016rsf}, allows for a more tractable analysis of the dynamics:
\iea{
  \mathcal{V}(\bar{\tilde{\sigma}},\tilde{\sigma})=\sum_j\left(\frac{2\lambda_j}{n_j}|\tilde{\sigma}_j(\phi_0)|^{n_j}+\frac{2\mu_j}{n_j'}|\tilde{\sigma}_j(\phi_0)|^{n_j'}\right),\label{eq:intkernel}
}
where the interaction couplings $\lambda_j$ and $\mu_j$ correspond to each mode $j$, assumed to satisfy the conditions $|\mu_j| \ll |\lambda_j| \ll |m_j^2|$. Additionally, we assume that $n_j' > n_j > 2$. While this approach is significantly simpler than full quantum geometric models, it still retains several key features that are relevant, capturing essential aspects of what might be expected from a universal effective behaviour. It is important to note that, at this stage, our effective action is not derived from a detailed underlying GFT model based on quantum geometric principles. Instead, we opt for the interaction kernel in equation \eqref{eq:intkernel}, as it is both manageable and shares similarities with certain microscopic GFT theories, such as those related to the EPRL model \cite{Oriti:2016qtz}. In this context, any GFT model capable of reproducing this effective action—either under the mean-field approximation or with some quantum corrections—would yield the same cosmic evolution that we explore in the following sections.

\paragraph{Equation of motion.} A crucial simplification arises from the structural absence of cross-coupling terms between distinct modes in the action. This diagonal structure induces a complete decoupling of equations of motion for different $j$-modes \cite{Oriti:2016qtz}, rendering the system analytically tractable. Notably, this structural feature manifests in prominent \lq fundamental models\rq\ of quantum gravity, including the EPRL spinfoam model and the Barrett-Crane formulation \cite{Jercher:2021bie} when restricted to their \textit{isotropic sector}. However, such mode decoupling constitutes a special property rather than a universal feature of GFTs. Generic GFT actions \cite{Baratin:2011hp} -- particularly in Riemannian signature formulations -- typically contain non-diagonal interaction terms that entangle different modes, substantially complicating both the perturbative analysis and non-perturbative treatment of the quantum theory.

The cosmological implications of monochromatic spin-mode interactions (where $j$-spectrum truncation to a single mode is imposed) were systematically investigated in \cite{deCesare:2016rsf}. Their analysis revealed that such Planck-scale interactions induce novel quantum gravitational corrections to effective universe dynamics. Particularly noteworthy is the emergence of a transient inflationary phase during primordial evolution -- a critical feature absent in purely classical FLRW cosmology. This inflationary mechanism arises naturally from the non-perturbative interference effects between discrete quantum geometry contributions, demonstrating how GFT condensate physics could bridge quantum gravity phenomenology with standard cosmological paradigms.

Our work extends this prior investigation of quantum gravitational condensates by implementing a multi-mode analysis that incorporates non-trivial $j$-spectrum contributions. Through this generalized framework, we demonstrate the emergence of late-time effective dark energy dynamics as a direct consequence of non-perturbative quantum gravitational effects -- without phenomenological insertion of exotic matter components. Crucially, while our model retains standard matter content (including necessary clock degrees of freedom) as required by relational quantization schemes, the observed cosmic acceleration arises purely from the underlying quantum gravity dynamics. This mechanism provides a concrete realization where Planck-scale quantum geometry imprints persist in the cosmological expansion history, challenging conventional dark energy paradigms through first-principles quantum gravity effects.

The dynamics are encoded in the action \eqref{eq:condaction}, and for later convenience we can transform the equation of motion into a more familiar hydrodynamic form\footnote{We will omit the subscript $0$ in $\phi$ for simplicity when there's no risk of confusion.}. 
Varying the action \eqref{eq:condaction} with respect to $\bar{\tilde{\sigma}}_j$ we get \cite{Marchetti:2020umh,deCesare:2016axk}
\iea{
  \tilde{\sigma}_j''-2\ii \tilde{\pi}_0\tilde{\sigma}_j-\xi_j^2\tilde{\sigma}_j+2\lambda_j|\tilde{\sigma}_j|^{n_j-2}\tilde{\sigma}_j+2\mu_j|\tilde{\sigma}_j|^{n_j'-2}\tilde{\sigma}_j=0.
}
Using the decomposition $\tilde{\sigma}_j(\phi) = \rho_j(\phi)e^{\ii \theta_j(\phi)}$ -- where $\rho_j\in\mathbb{R}^+$ represents the condensate number density and $\theta_j$ its quantum phase -- transforms the complex dynamical equations into coupled hydrodynamic relations. The global $\mathrm{U}(1)$ symmetry $\theta_j \to \theta_j + \alpha$ in the effective action generates a conserved Noether current, and the corresponding conserved charge reads \cite{Oriti:2016qtz,Marchetti:2020umh}
\iea{
  Q_j=(\theta_j'-\tilde{\pi}_0)\rho_j^2. 
}
Using the definition of $Q_j$ we can write the module equation as following~\cite{deCesare:2016rsf,Marchetti:2020umh}
\iea{
  \rho_j''-\frac{Q_j^2}{\rho_j^3}-m_j^2\rho_j+\lambda_j\rho_j^{n_j-1}+\mu_j\rho_j^{n_j'-1}=0. \label{eq:rhojpp}
}
In this equation, we have introduced a new constant $\displaystyle m_j^2=\xi_j^2-\tilde{\pi}_0^2$. By integrating the equation once, we can get another conserved quantity \cite{Oriti:2016qtz}, due to the invariance under \lq\lq clock-time translation\rq\rq\  \cite{deCesare:2016rsf,Marchetti:2020umh},
\iea{
  E_j=\frac{1}{2}(\rho_j')^2-\frac{1}{2}m_j^2\rho_j^2+\frac{Q_j^2}{2\rho_j^2}+\frac{\lambda_j}{n_j}\rho_j^{n_j}+\frac{\mu_j}{n_j'}\rho_j^{n_j'}. \label{eq:defej}
}

\subsection{Volume dynamics} \label{sec:volumedynamics}

We have mentioned that the universe evolution can be seen from the volume, in particular, for our condensate state $\ket{\sigma}$, the expectation value of the volume operator $\hat{V}$ can be calculated as following \cite{Oriti:2016qtz, Marchetti:2020qsq,Oriti:2021rvm}
\iea{
  V(\phi_0)&=&\braket{\sigma_\varepsilon;\phi_0,\pi_0|\hat{V}|\sigma_\varepsilon;\phi_0,\pi_0} \nonumber \\
  &=&\braket{\sigma_\varepsilon;\phi_0,\pi_0|\int\dd\phi\sum_{\vec{x},\vec{x}'}V(\iota,\iota')\delta_{\vec{x}-\{\iota\},\vec{x}'-\{\iota'\}}\hat{c}_{\vec{x}}^\dag(\phi)\hat{c}_{\vec{x}'}(\phi)|\sigma_\varepsilon;\phi_0,\pi_0} \nonumber \\
  &\approx&\sum_jV_j\rho_j(\phi_0)^2. \label{eq:totalVcondensate}
}
The volume contribution from each quantum (tetrahedron) in the spin $j$ representation is given by $V_j \propto l_p^3 j^{3/2}$, where $l_p$ denotes the Planck length. We also employ the intertwiner normalization condition $\sum_{\vec{m}}\mathcal{I}^{j,\iota_+}_{\vec{m}} \bar{\mathcal{I}}^{j,\iota'_+}_{\vec{m}} = \delta_{\iota,\iota'}$. The approximation used in this analysis consists of retaining only the dominant contribution to the saddle-point approximation of the peaking function, which arises from our choice of state. To avoid confusion, we reintroduce the subscript $0$ for the given relational time $\phi_0$ at this stage \cite{Marchetti:2020umh}.

Having derived the equation of motion \eqref{eq:rhojpp} of condensate density $\rho_j$, we can recast the volume dynamics in the form of modified FLRW equations \cite{Oriti:2016qtz}
\iea{
  \left(\frac{V'}{3V}\right)^2&=&\left[\frac{2\sum_j V_j\sqrt{2E_j\rho_j^2-Q_j^2+m_j^2\rho_j^4-\frac{2}{n_j}\lambda_j\rho_j^{n_j+2}-\frac{2}{n'_j}\mu_j\rho_j^{n'_j+2}}}{3\sum_k V_k\rho_k^2}\right]^2, \label{eq:vpsquare}\\
  \frac{V''}{V}&=&\frac{2\sum_j V_j\left[2E_j+2m_j^2\rho_j^2-\left(1+\frac{2}{n_j}\right)\lambda_j\rho_j^{n_j}-\left(1+\frac{2}{n_j'}\right)\mu_j\rho_j^{n_j'}\right]}{\sum_k V_k\rho_k^2}, \label{eq:vpp}
}
where the first order equation \eqref{eq:defej} is also used, and since only the expansion phase will be considered in the current work, we have chosen the positive root of $\rho_j'$ in equation \eqref{eq:defej}.

\paragraph{Classical regime.} 
During the pre-interaction dominated phase of volumetric expansion, the system transitions through an intermediate regime where the dynamics approximate FLRW cosmology with a free massless scalar field \cite{Oriti:2016qtz,Marchetti:2020umh}. This occurs under the hierarchy of scales: $\rho_j^2 \gg E_j/m_j^2, \quad \rho_j^3 \gg Q_j^2/m_j^2,~|\mu_j|\rho_j^{n_j'-2} \ll |\lambda_j|\rho_j^{n_j-2} \ll m_j^2$, under which the fundamental equations \eqref{eq:vpsquare} and \eqref{eq:vpp} reduce to
\ieas{
 \left(\frac{V'}{3V}\right)^2&=&\left(\frac{2\sum_j V_jm_j\rho_j^2}{3\sum_k V_k\rho_k^2}\right)^2, ~
  \frac{V''}{V}=\frac{\sum_j V_j\left(4m_j^2\rho_j^2\right)}{\sum_k V_k\rho_k^2}. 
}
For a dominant spin mode $\tilde{j}$ with approximately constant $m_{\tilde{j}}$\footnote{A sufficient but not necessary condition for FLWR emergence}, we identify:
\begin{equation*}
    m_{\tilde{j}}^2 \equiv 3\pi G \quad .
\end{equation*}
where $G$ represents the effective dimensionless Newton constant. This yields the characteristic FLRW equations in relational time:
\ieas{
  \left(\frac{V'}{V}\right)^2=\frac{V''}{V}=12\pi G.
}
Notably, mode dominance dynamics \cite{Gielen:2016uft} ensure rapid convergence to the lowest spin mode $j_0$, making $m_{j_0}^2 = 3\pi G$ sufficient for classical recovery. This establishes GFT condensate hydrodynamics as a viable pathway to classical cosmology at macroscopic scales, provided interaction terms remain subdominant. Furthermore, single-mode dominance further induces a direct proportionality between the Hubble parameter and volume evolution rate \cite{deCesare:2016rsf,Oriti:2021rvm}:
\begin{IEEEeqnarray}{rCl}
	H &=& \frac{Q_1}{3}\frac{\mathcal{V}'}{\mathcal{V}} . \label{eq:HVrelation}
\end{IEEEeqnarray} 
This critical relationship enables the extraction of the cosmological constant from microscopic GFT parameters, as demonstrated in subsequent analysis.

The central question, therefore, is how the interactions within GFT influence the effective dynamics. This is the primary focus of the present work, which builds upon the initial analyses conducted in \cite{Pithis:2016cxg,Pithis:2016wzf,deCesare:2016rsf} and further develops the framework introduced in \cite{Oriti:2021rvm}.

\section{Effective equation of state and its dynamics} \label{sec:effwsingle}
To characterize effective cosmological dynamics, one can model them as an effective matter component defined entirely by its equation of state.

In a homogeneous universe, the matter content is modeled as a perfect fluid characterized by energy density $\rho$ and pressure $p$, related by the equation of state $w = p/\rho$. This determines cosmological evolution; for instance, $w < -1/3$ leads to accelerated expansion. Observations indicate $w \simeq -1$, consistent with cosmic acceleration, whereas standard matter corresponds to $w = 1/3$ (relativistic) or $w = 0$ (non-relativistic). While a cosmological constant could reproduce this value, it raises questions about its origin, quantum dynamics, interactions with quantum gravity, and temporal variation---central aspects of the \emph{dark energy problem} \cite{Martin:2012bt}.

We now recast our emergent cosmological dynamics using this framework, expressed via relational clock evolution.

For a homogeneous, isotropic metric with scale factor $a(t)$, the Hubble parameter is $H = \dot{a}/a$, where the dot denotes a comoving time derivative. The effective equation of state becomes $w = -1 - 2\dot{H}/(3H^2)$. Using the relational clock (see section~\ref{sec:gftcosreview}), it takes the form \cite{Oriti:2021rvm}
\iea{
  w = 3 - \frac{2 V V''}{(V')^2}, \label{eq:effwdef}
}
where $V = a^3$ is the volume and primes denote derivatives with respect to relational time $\phi$.

\subsection{$w$ from single-mode GFT condensates}
Before presenting our new analysis of interacting GFT cosmologies, we contextualize key findings from previous work. The previous study \cite{deCesare:2016rsf} analyzed effective equation of state behavior under single-mode dominance, using the same mean-field approximation employed here. While multi-mode scenarios introduce richer transitional dynamics and modified convergence to asymptotic $w$ values, this single-mode framework already reveals crucial insights into how GFT interactions shape emergent cosmology.

For a single $j$ mode, combining equations \eqref{eq:vpsquare}, \eqref{eq:vpp}, and \eqref{eq:effwdef} yields:
\iea{
  w = \frac{-3Q^2 + 4E\rho^2 + m^2\rho^4 + \left(1-\frac{4}{n}\right)\lambda\rho^{n+2} + \left(1-\frac{4}{n'}\right)\mu\rho^{n'+2}}{-Q^2 + 2E\rho^2 + m^2\rho^4 - \frac{2}{n}\lambda\rho^{n+2} - \frac{2}{n'}\mu\rho^{n'+2}},
}
where mode indices are omitted for simplicity. The direct proportionality $V \propto \rho^2$ enables $w(V)$ analysis without solving equations of motion - a critical simplification facilitating analytical progress.

\paragraph{Bounce.} In the interaction-free limit ($\lambda=\mu=0$) at small condensate modulus $\rho$, the equation of state reduces to:
\ieas{
  w = \frac{-3Q^2 + 4E\rho^2 + m^2\rho^4}{-Q^2 + 2E\rho^2 + m^2\rho^4}.
}
The bounce condition $-Q^2 + 2E\rho^2 + m\rho^4 = 0$ determines the minimum volume:
\ieas{
  \rho_b = \frac{1}{m}\sqrt{\sqrt{E^2 + m^2Q^2} - E}.
}
At $\rho_b$, the numerator becomes negative, driving $w \to -\infty$ - characteristic of post-bounce acceleration\footnote{Acceleration follows from $V''_b > 0$ at the bounce ($V'_b=0$), consistent with \eqref{eq:effwdef} behavior as $V \to V_b$.}. Crucially, this inflationary phase terminates rapidly, with acceleration ending at volumes comparable to $V_b$ \cite{deCesare:2016rsf}. {And we can see that when there's a bounce, i.e., when $\rho_b\neq 0$, we must have $Q\neq 0$. More generally, we will get a bounce if at least one $Q_j$ is non-vanishing in all condensate modes~\cite{Gielen:2016uft,Oriti:2021rvm}.}

\paragraph{Phantom divide crossing.} Single-mode dynamics permit limited phantom behavior:
\begin{itemize}
    \item \textbf{$\mu<0$ regime:} Asymptotic $w$ approaches $2-n'/2$ with:
    \ieas{
      w \to -1 \quad (n'=6) \quad \text{or} \quad w < -1 \quad (n'>6);
    }
    \item \textbf{Late time singularities:} $n'>6$ creates divergent $\rho_\psi$ lead to Big Rip \cite{Caldwell:2003vq};
    \item \textbf{Multi-mode resolution:} Two-mode systems enable phantom crossing ($w<-1$) while avoiding singularities through phantom de Sitter analogs \cite{McInnes:2001zw,Oriti:2021rvm};
\end{itemize}
The critical distinction lies in mode competition - multiple modes regulate energy density divergence despite phantom behavior, as detailed in subsequent analysis.

\subsection{$w$ from the two-modes \acrshort{gft} condensates}
The analysis above has been extended to mean fields depending on two modes in \cite{Oriti:2021rvm}. We now provide a summary of the relevant results from this analysis before discussing them in more detail in the next section.

\paragraph{The interacting case with a single coupling $\lambda_j$.} In the presence of interactions, it becomes much harder to find an exact solution of the equation of motion \eqref{eq:defej}, with two modes contributing. Still, an approximate solution in the large volume region can be obtained. In fact, when $\rho$ is large, we can ignore $E_j,~Q_j$ and $m_j$ terms in the equation of motion \eqref{eq:defej}, then we have approximately\footnote{We have chosen $\mu_j=0,~\lambda_j<0$ and $n_j=6$, which are responsible for the emergence of the late time de Sitter acceleration with \acrshort{eos} $w=-1$.}
\begin{IEEEeqnarray}{rCl}
  \rho_j'(\phi)=\sqrt{-\frac{\lambda_j}{3}}\rho_j(\phi)^{3}, 
\end{IEEEeqnarray}
which can be solved to give~\cite{Oriti:2021rvm}
\begin{IEEEeqnarray}{rCl}
	\rho_{j}(\phi)=\frac{3^{1/4}}{\sqrt{2\sqrt{-\lambda_j}\left(\phi_{j\infty}-\phi\right)}}. \label{eq:largerhosol}
\end{IEEEeqnarray} 
with $\phi_{j\infty}$ is a constant. 
We see that when interactions are included, the relational time $\phi$ will have an upper bound, and the mode associated with the smallest asymptotic value $\phi_{j\infty}$ governs the late-time dynamics. This leads to an asymptotic de Sitter-like phase. At intermediate volumes, contributions from other modes remain significant, collectively driving the universe's expansion. As demonstrated in \cite{Oriti:2021rvm}, for vanishing high-order interaction couplings ($\mu_1 = \mu_2 = 0$) and small $\lambda_1$, $\lambda_2$, the equation of state parameter $w$ is dominated by the free sector of the condensate at small volumes. In this regime, $w$ approaches $w = 1$ from below as the volume increases, recovering the FLRW universe of standard cosmology. However, at larger volumes, interaction terms progressively dictate the condensate dynamics. When the scale parameters $\rho_j$ (and consequently the volume) grow sufficiently large, $w$ becomes dominated by these interaction terms. For the specific case $n_1 = n_2 = 6$ (defining the interaction order), $w$ in this interaction-dominated regime depends solely on the ratio $r = \rho_2/\rho_1$, yielding
\iea{
  w  &=& -1-\frac{4\mathcal{V}_1\mathcal{V}_2r^2\left(r^{2}-\sqrt{\lambda_1/\lambda_2}\right)^2}{\left(\sqrt{\lambda_1/\lambda_2}\mathcal{V}_1+\mathcal{V}_2r^{4}\right)^2} \qquad . \label{eq:wrrho2rho1}
}
Since the parameters are all real and both couplings $\lambda_1$ and $\lambda_2$ are assumed to be negative, we see that $\displaystyle w\leq-1$. Recall that when the volume is large, one of the two modes will dominate over the other, and then we have $r\to0$ or $r\to\infty$. In either case, $w$ will approach $\displaystyle -1$ from below, in contrast with the single mode case discussed previously.

When the universe volume gets larger, the ratio $r$ gets smaller and we can expand $w$ respect to small $r$, such that the \acrshort{eos} can be expressed by only the total volume $\mathcal{V}$ as following
\begin{IEEEeqnarray}{rCl}
	w &=& -1-\frac{b}{\mathcal{V}} , \label{eq:wvappro}
\end{IEEEeqnarray} 
where $b=4\mathcal{V}_2\rho_{2}^2(\phi_{1\infty}){>0}$ is a { positive} constant.

{Before proceeding, it is worth discussing the effects of including more than two condensate modes. Let $\phi_{1\infty}$ denote the smallest asymptotic value of the relational time for mode $\rho_1$. At very late times, the evolution is therefore governed by this mode, according to the asymptotic solution~\eqref{eq:largerhosol}. Introducing an additional mode with $\phi_{2\infty} > \phi_{1\infty}$ modifies the evolution by inducing a phantom phase before the onset of the de Sitter expansion, without altering the asymptotic behavior. If a third mode $\rho_3$ is included, we must require that $\phi_{3\infty} > \phi_{2\infty}$; otherwise, the $\rho_3$ mode would dominate the phantom phase. Consequently, we expect that the inclusion of additional modes primarily affects the cosmological evolution before the phantom phase and should not significantly alter the main results presented in the current work.}

\section{Detailed behaviour of the effective \acrshort{eos} in the two-mode case}
\label{sec:effwbehaviour}
In this section, we present several further results on the cosmological evolution, in the same phenomenological model of quantum gravity condensates, in the two-mode case. These include: a) an estimate of the location of the phantom crossing, and b) the explicit expression of the asymptotic cosmological as a function of the parameters of the model and the condensate state underlying it.

\subsection{Location of phantom crossing}
To determine the location of the phantom crossing along the cosmic history, we adopt an indirect strategy. First, we argue that the position of phantom crossing is close to the minimal point of $w$. Second, we identify the minimal point of $w$ by solving the equation $w'=0$, using the definition of $w$. Third, we solve the equation $w'=0$ approximately based on the large volume solution and the assumption that at the minima the value of $\phi$ is close to $\phi_1$, where the volume diverges. 

The necessity of such an indirect strategy stems from the fact that, to get the exact position of the phantom crossing, one should also consider the free terms and the resulting equation is hard to solve analytically. 

To be more precise, in principle, by substituting the solution \eqref{eq:largerhosol} of $\rho_j$ into total volume \eqref{eq:totalVcondensate}, and then to the effective \acrshort{eos} \eqref{eq:effwdef}, we can obtain $w=w(\phi)$. Solving $w(\phi)=-1$ will provide us with the position of the phantom crossing. However, in obtaining the solution \eqref{eq:largerhosol}, we used the large volume approximation by ignoring all the free terms, which results in an \acrshort{eos} always less than $-1$ \cite{Oriti:2021rvm}, as shown by the approximation \eqref{eq:wrrho2rho1}.
Therefore, to get a meaningful equation $w(\phi)=-1$ we need to use the exact form of $w$. This requires exact solutions of the equation of motion \eqref{eq:defej}, which is quite difficult as equation \eqref{eq:defej} is non-linear.

One can work around this by noting that the position of the phantom crossing must lie between the end of the Friedman phase (where $w\simeq 1$) and the position where $w$ reaches its minimal value. We now show that the latter two positions are close to each other.%
 Furthermore, under the large volume approximation we are using, we can determine the minimum value of $w$.

\paragraph{Minimal value of $w$.} We need to solve the equation $w'=0$. Taking derivative respect to relational time $\phi$ on both sides of effective \acrshort{eos} \eqref{eq:effwdef}, we obtain
\begin{IEEEeqnarray}{rCl}
	w'=-\frac{2\left[\left(\mathcal{V}'\right)^2\mathcal{V}''-2\mathcal{V}\left(\mathcal{V}''\right)^2+\mathcal{V}  \mathcal{V}' \mathcal{V}'''\right]}{\left(\mathcal{V}'\right)^3}. \label{eq:wpdef}
\end{IEEEeqnarray} 
In the two-modes case, the total volume has the form  
\begin{IEEEeqnarray*}{rCl}
  \mathcal{V}(\phi)=\mathcal{V}_1\rho_1^2(\phi)+\mathcal{V}_2\rho_2^2(\phi).
\end{IEEEeqnarray*}
When the volume is large, substituting the total volume and the corresponding solutions \eqref{eq:largerhosol} into equation \eqref{eq:wpdef}, we obtain
\begin{align} %
	w'&= \frac{4\mathcal{V}_1\mathcal{V}_2(\lambda_1\lambda_2)^{\frac{3}{2}}\left(\phi_{2\infty}-\phi_{1\infty}\right)^2}{\left[\mathcal{V}_2\lambda_1\sqrt{-\lambda_2}\left(\phi_{1\infty}-\phi\right)^2+\mathcal{V}_1\sqrt{-\lambda_1}\lambda_2\left(\phi_{2\infty}-\phi\right)^2\right]^3} \nonumber  \\ &\times\left[\mathcal{V}_2\lambda_1\sqrt{-\lambda_2}\left(\phi_{1\infty}-\phi\right)^2\left(2\phi+\phi_{1\infty}-3\phi_{2\infty}\right)+\mathcal{V}_1\sqrt{-\lambda_1}\lambda_2\left(\phi_{2\infty}-\phi\right)^2\left(2\phi-3\phi_{1\infty}+\phi_{2\infty}\right)\right]. \label{eq:wpsub}
\end{align} 
We see that if $\phi_{1\infty}=\phi_{2\infty}$ then $w'=0$ as we expected. For $\phi_{1\infty}\neq\phi_{2\infty}$, demanding $w'=0$ is equivalent to require the second line of \eqref{eq:wpsub} to vanish, which provides us a cubic equation of $\phi$
\begin{align}
\mathcal{V}_2\lambda_1\sqrt{-\lambda_2}\left(\phi_{1\infty}-\phi\right)^2\left(2\phi+\phi_{1\infty}-3\phi_{2\infty}\right) +\mathcal{V}_1\sqrt{-\lambda_1}\lambda_2\left(\phi_{2\infty}-\phi\right)^2\left(2\phi-3\phi_{1\infty}+\phi_{2\infty}\right)=0.%
\end{align}
In principle, this equation can be solved exactly, but the exact solution is not that useful, and for now it is sufficient to consider approximate solutions. In fact, at the minima the value of $\phi$ is close to $\phi_{1\infty}$, so we can assume that $\phi=\phi_{1\infty}-\delta$ and then expand $w'$ with respect to $\delta$. At leading order, we have
\begin{IEEEeqnarray}{rCl}
  w'=\frac{4\mathcal{V}_2\lambda_1\left(6\delta+\phi_{1\infty}-\phi_{2\infty}\right)}{\mathcal{V}_1\sqrt{\lambda_1\lambda_2}(\phi_{2\infty}-\phi_{1\infty})^2}.
\end{IEEEeqnarray}
Solving $w'=0$ gives
\begin{IEEEeqnarray}{rCl}
  \delta=\frac{\phi_{2\infty}-\phi_{1\infty}}{6}, \label{eq:wp0delta}
\end{IEEEeqnarray}
which results in 
\begin{IEEEeqnarray}{rCl}
  \mathcal{V}_{\mathrm{min},w}&=& -\frac{3\sqrt{3}\left(\mathcal{V}_2\lambda_1\sqrt{-\lambda_2}+7\mathcal{V}_1\sqrt{-\lambda_1}\lambda_2\right)}{7\lambda_1\lambda_2\left(\phi_{2\infty}-\phi_{1\infty}\right)}=\frac{3\sqrt{3}}{7(\phi_{2\infty}-\phi_{1\infty})}\left(\frac{\mathcal{V}_2}{\sqrt{-\lambda_2}}+\frac{7\mathcal{V}_1}{\sqrt{-\lambda_1}}\right), \nonumber \\ \label{eq:Vatminima}\\
  w_{\mathrm{min}}&=& \frac{\mathcal{V}_2^2\lambda_1+2401\mathcal{V}_1^2\lambda_2-1106\mathcal{V}_1\mathcal{V}_2\sqrt{\lambda_1\lambda_2}}{\left(\mathcal{V}_2\sqrt{-\lambda_1}+49\mathcal{V}_1\sqrt{-\lambda_2}\right)^2} = -1-\frac{1008\mathcal{V}_1\mathcal{V}_2\sqrt{\lambda_1\lambda_2}}{\left(\mathcal{V}_2\sqrt{-\lambda_1}+49\mathcal{V}_1\sqrt{-\lambda_2}\right)^2}. \nonumber \\ \label{eq:wminimum}
\end{IEEEeqnarray}

Note that $\mathcal{V}_{\mathrm{min},w}$ is of the same order as $1/\sqrt{\lambda_j}$, around which the mass term $m_j^2\rho_j^2$ and interaction term $\lambda_j\rho_j^6/3$ in the equation of motion \eqref{eq:vpsquare} are of the same order. This indicates the end of the Friedman phase. Furthermore, the phantom crossing, where the \acrshort{eos} $w$ starts to become less than $-1$, should happen before $w$ reaches its minimum. Since $\mathcal{V}_{\mathrm{min},w}$ is close to the end of the Friedman phase, we see that the universe becomes phantom-like soon after the Friedman phase ends, and $w$ will reach its minimum soon after phantom crossing. 

\begin{figure}[htpb]
	\centering
	\includegraphics[width=0.6\textwidth]{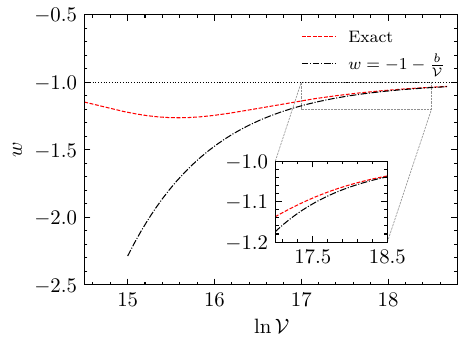}
	\caption{The comparison between the exact value and approximate value of \acrshort{eos} $w$. Parameters are $\mathcal{V}_1=1/3,m_1^2=3,~E_1=5,~Q_1^2=9,\mathcal{V}_2=1/2,~m_2^2=2,~E_2=9,~Q_2^2=2.25,~\lambda_1=-10^{-8}, ~\lambda_2=-9.5\times10^{-8}$.}
	\label{fig:wlnvpara38exaappcomp}
\end{figure}

\paragraph{Beyond the minimum.} On the other hand, as shown in figure \ref{fig:wlnvpara38exaappcomp}, the approximation \eqref{eq:wvappro} becomes accurate soon after $w$ reaches its minimum value. In fact,  we can substitute $\mathcal{V}_{\mathrm{min},w}$ into our approximation \eqref{eq:wvappro}, and obtain 
\begin{IEEEeqnarray}{rCl}
	w &=& -1-\frac{56\mathcal{V}_2\sqrt{-\lambda_1}}{3\mathcal{V}_2\sqrt{-\lambda_1}+21\mathcal{V}_1\sqrt{-\lambda_2}} = -1 - \frac{56\mathcal{V}_1\mathcal{V}_2\sqrt{\lambda_1\lambda_2}}{3\mathcal{V}_1\mathcal{V}_2\sqrt{\lambda_1\lambda_2}-21\mathcal{V}_1^2\lambda_2}. 
\end{IEEEeqnarray} 
This is close to the minimal value \eqref{eq:wminimum}, which means that $w$ can be approximated using \eqref{eq:wvappro} soon after $w$ passes its minimum value. Furthermore, the approximation \eqref{eq:wvappro} approaches $-1$ quickly, hence in the phantom phase, $w$ can only deviate from $-1$ significantly near its minimum.

{In figure~\ref{fig:wlnvpara404142} (see appendix \ref{app:otherpara}), we numerically show the evolution of the \acrshort{eos} parameter $w$ for three sets of parameters. We observe that the minimum value of $w$ indeed occurs near the point where it crosses the phantom divide at $w = -1$. However, to demonstrate this result more generally for other values of parameters, one must rely on the approximation used above.
}

We conclude that after phantom crossing, $w$ will approach its asymptotic value $-1$ quickly in our model. Therefore, if we are experiencing a phantom phase, the phantom crossing must have happened recently. We emphasize that this is not trivial, as our model unifies two observational facts:
\begin{enumerate}
	\item the phantom crossing happened recently, and
	\item $w$ deviates from $-1$ notably.
\end{enumerate}
Our claim that these two facts are related in our model: $w$ may have a notable deviation from $-1$ if and only if the phantom phase has happened recently (at a low redshift).

\subsection{Explicit form of cosmological constant}
In our GFT model, a de Sitter phase with an effective cosmological constant emerges asymptotically from the microscopic quantum dynamics of spacetime constituents. We can determine the precise expression of such an effective cosmological constant as a function of the microscopic parameters of the model and of the underlying quantum state of the universe. 

To do so, note that in the asymptotic large volume region, we can ignore the contributions from other terms and only keep (order-6) interactions in the equation of motion \eqref{eq:vpsquare}. Furthermore, we have seen that in such a region only a single mode dominates~\cite{Oriti:2021rvm}, and we can use the relation \eqref{eq:HVrelation} between Hubble parameter $H$ and the ratio $\mathcal{V}'/\mathcal{V}$. Therefore, in the asymptotic de Sitter regime we obtain
\begin{IEEEeqnarray}{rCl}
  H^2&=& \frac{8}{9}Q_1^2\left(\frac{-\lambda_1}{6V_1^2}\right)=\frac{1}{3}\left[\frac{4Q_1^2}{3V_1^2}(-\lambda_1)\right].
\end{IEEEeqnarray} 
Comparing this equation with the $\Lambda$CDM model at late time, we see that the cosmological constant is determined by the microscopic parameters of \acrshort{gft} model~\cite{deCesare:2016rsf}
\begin{IEEEeqnarray}{rCl}
	\Lambda=\frac{4Q_1^2}{3V_1^2}(-\lambda_1). \label{eq:Lambdaqv}
\end{IEEEeqnarray}
Let us discuss the implications of this expression for $\Lambda$, and emphasise several points of interest.
\begin{itemize}
	\item $\Lambda$ is determined by the parameters of a single mode despite the fact that two modes are considered; 
        \item for a non-vanishing $\Lambda$, we see that $Q_1\neq 0$ is necessary; 
        {From the theory \emph{per se}, one can in principle consider the case where $Q_j$ vanishes for all modes $j$~\cite{Marchetti:2020umh}\footnote{Cf equation (5.15) and section 5.5 in~\cite{Marchetti:2020umh}.}, which may provide us the Minkowski spacetime~\cite{Oriti:2016qtz}. But equation \eqref{eq:Lambdaqv} shows that,}
        the observation of a non-vanishing cosmological constant at late times would require that the $Q_1$ can't be zero, in turn implying a quantum bounce~\cite{Oriti:2016qtz}, which resolves the Big Bang singularity. In other words, the non-vanishing cosmological constant $\Lambda$ itself would be a remnant of the expansion history of our universe in the far past, similar to the CMB (although of course of an entirely different kind); this indirect connection between very early and very late universe dynamics (and thus very small and very large scales) is quite intriguing, and only possible because we are in an emergent spacetime framework. %
        
        {In general, the value of $\Lambda$ cannot be directly related to the bouncing scale, since $\Lambda$ is determined by the parameters of a single mode, whereas the bounce depends on the collective contributions from all modes. This is because, in the bouncing regime, the modulus $\rho_j$ of each mode remains small, and no single mode dominates the dynamics~\cite{Oriti:2021rvm}. This is actually required, as to ensure that the scalar field $\phi$ serves as a well-behaved relational clock, the sum of all $Q_j$ must vanish~\cite{Marchetti:2020umh}. If instead a single mode $\rho_1$ were to dominate the dynamics from the beginning of the bounce, then enforcing $\phi$ as a good clock would necessitate $Q_1 = 0$, leading to a vanishing cosmological constant. By contrast, when multiple modes contribute to the bounce while a single mode dominates the late-time dynamics, it becomes possible to satisfy $\sum_j Q_j = 0$ and still retain a nonzero $Q_1$, thereby allowing for a nonvanishing cosmological constant.
        
        Nevertheless, we can extract potentially relevant observational constraints, at least qualitatively.}
        On the one hand, $Q_1$ can't be too large, otherwise, we will have a large cosmological constant $\Lambda$; on the other hand, $Q_1$ contributes to the critical energy density of the universe at the bounce \cite{Oriti:2016qtz}, hence it can't be too small, or we would lose the established physics of the hot dense state of the universe in the very early time, with spacetime dynamics still governed by the Friedmann evolution, since the universe would enter instead quickly in a quantum gravity bouncing regime. In principle, these two constrain can be used to narrow down the possible range of $Q_1$ using observations (once we implement the more realistic matter contents into our \acrshort{gft} formalism).
	\item $\Lambda$ does not depend on $m_j$, hence the mass renormalization of \acrshort{gft} model will not change the value of cosmological constant; note also that $m_j$ is related to the effective Newton's constant, as emerging in the Friedmann phase of cosmic expansion; the two key couplings of gravitational dynamics in classical GR are thus both emergent and independent of each other in this quantum gravity model; 
	\item $V_1$ is the volume of a spacetime quantum in the condensate state governed by the mean-field hydrodynamic equations we have recast as cosmological dynamics; it is also the minimal volume that the universe can reach, since the dynamics tend to favour the dominance of the minimal spin value $j$ at late times; thus, the cosmological constant carries another signature of the deep quantum gravity physics; this value, moreover, is not affected by renormalization;  
    \item Similarly, $Q_1$ is an integral constant from the equation of motion, characterizing the quantum state of the universe, and it is also unaffected by renormalization; they both remain the same under renormalization. 
    \item on the other hand, $\Lambda$ depends on the coupling of the order-6 interaction, whose renormalization will therefore determine the running of the effective cosmological constant. %
    {A complete RG analysis of quantum geometric TGFT models for $4$d quantum gravity is missing, thus we cannot give definite statements about how the underlying RG flow affects the effective cosmological constant. However, the analysis performed within a Landau-Ginzburg approach in \cite{Marchetti:2022nrf,Marchetti:2022igl} gives some interesting hints. It suggests that, at a mean field level, a condensate phase can indeed be produced and that the relevant critical point at low RG scale, which corresponds to a small-$j$ regime, is a Gaussian one, thus one of vanishing interaction couplings. This is consistent with our mean field analysis of the cosmological dynamics, which is indeed restricted to small-$j$ modes, and it would suggest that in fact the effective cosmological constant would be driven towards a vanishing value by the RG flow, even if it has a non-vanishing bare value.  } 
\end{itemize}

\section{The deviation of Hubble parameter $H_0$ compared to single mode case} 
\label{sec:deviationsingle}

In this section, we perform a more quantitative analysis of the way the presence of a second dynamical spin mode increases the value of $H_0$, with respect to the one determined by a single mode, and thus by a non-dynamical cosmological constant. We hope this represents a step toward a direct comparison of our emergent cosmological dynamics with observational data, and hints at a possible quantum gravity resolution of the $H_0$-tension ~\cite{Yang:2021hxg}.

The method we follow is based on the one employed in \cite{Heisenberg:2022gqk}. %
The key steps are: 1. to write down the single mode Friedmann equation, assuming that the preferred value of Hubble parameter is $H_0$ when we fit the cosmological data; 2. to express the effect of introducing a second mode (akin to adding a phantom-like contribution to the Friedmann equation (equation \eqref{eq:Hubbletwomodes})) as a small modification $\delta H$ of this base value; 3. to relate the deviation $\delta H_0$ of the current Hubble parameter and $\delta H$ by the response function $\mathcal{R}_{H_0}$, which can be then computed (section \ref{sec:responsefuctionRH0}); we also show that the regions where $\delta H$ and $\mathcal{R}_{H_0}$ deviate from $0$ overlap, so $\delta H_0$ will indeed change (equation \eqref{eq:deltaH0RH0}); we do this by showing that their minimal points are close.
Again, this indirect procedure allows us to see the effects of a second mode on the observed quantities, without solving the full evolution equation in the two-modes dynamics.

{We emphasise again that, in our model, radiation and non-relativistic matter are not included, and thus a direct comparison of our GFT-derived cosmological dynamics with observations is not possible. In this work, we try to achieve a less ambitious goal, that is to show that the inclusion of a second mode will, rather generically, increase the value of Hubble parameter $H_0$ compared to the single mode case.}

In the previous section, we have seen that the phantom crossing can only occur recently, which enables us to regard the two-modes evolution as a slight modification of the single-mode case at late times. Such modifications will change the values of the parameters in our model, for example, the presence of phantom phase will increase the current Hubble parameter $H_0$ inferred from \acrshort{cmb} data~\cite{Heisenberg:2022gqk}. Since the parameters in the single mode case are easier to fix using the observational data, the cosmological quantities in the presence of a second mode can be deduced approximately. In particular, the Hubble parameter could be expressed as $H_0+\delta H_0$, where $H_0$ is the value deduced from the single mode evolution and $\delta H_0$ is a small deviation.

For a single mode, we can substitute equation \eqref{eq:vpsquare} into equation \eqref{eq:HVrelation} to obtain~\cite{deCesare:2016rsf}, 
\begin{IEEEeqnarray}{rCl}
	H_{\mathrm{single}}^2 &=&\frac{8Q_1^2}{9}\left(\frac{\varepsilon_{Q_1}}{\mathcal{V}^4}+\frac{\varepsilon_{E_1}}{\mathcal{V}^3}+\frac{\varepsilon_{m_1}}{\mathcal{V}^2}+\varepsilon_{\lambda_1}\right)  ,  \label{eq:FLRWHsingle}
\end{IEEEeqnarray} 
with the parameters
\begin{IEEEeqnarray*}{rCl}
	\varepsilon_{Q_1} &=& -\frac{Q_1^2}{2}\mathcal{V}_1^2 , \quad
	\varepsilon_{E_1} = \mathcal{V}_1E_1 ,  \quad
	\varepsilon_{m_1} = \frac{m_1^2}{2} ,  \quad
	\varepsilon_{\lambda_1} = -\frac{\lambda_1}{6\mathcal{V}_1^2} .
\end{IEEEeqnarray*} 
In terms of redshift $z$ such that $(1+z)^3=\mathcal{V}_0/\mathcal{V}$ for the current volume $\mathcal{V}_0$, equation \eqref{eq:FLRWHsingle} can be written as 
\begin{IEEEeqnarray}{rCl}
	H_{\mathrm{single}}^2 &=& H_0^2 \left[\Omega_{Q_1}(1+z)^{12}+\Omega_{E_1}(1+z)^9+\Omega_{m_1}(1+z)^6+\Omega_{\lambda_1}\right], 
\end{IEEEeqnarray} 
where we defined
\begin{IEEEeqnarray*}{rCl}
	H_0^2 &=& \frac{8Q_1^2}{9}\left(\frac{\varepsilon_{Q_1}}{\mathcal{V}_0^4}+\frac{\varepsilon_{E_1}}{\mathcal{V}_0^3}+\frac{\varepsilon_{m_1}}{\mathcal{V}_0^2}+\varepsilon_{\lambda_1}\right) , \\
	\Omega_{Q_1} &=& \frac{1}{H_0^2} \frac{8Q_1^2}{9}\frac{\varepsilon_{Q_1}}{\mathcal{V}_0^4},  \quad \Omega_{E_1} = \frac{1}{H_0^2} \frac{8Q_1^2}{9}\frac{\varepsilon_{E_1}}{\mathcal{V}_0^3} ,  \\
	\Omega_{m_1} &=& \frac{1}{H_0^2} \frac{8Q_1^2}{9}\frac{\varepsilon_{m_1}}{\mathcal{V}_0^2} ,  \quad\Omega_{\lambda_1} = \frac{1}{H_0^2} \frac{8Q_1^2}{9}\varepsilon_{\lambda_1},  
\end{IEEEeqnarray*} 
such that $\Omega_{Q_1}+\Omega_{E_1}+\Omega_{m_1}+\Omega_{\lambda_1}=1$, and the value of $H_0$ can be determined by fitting observed data. Since the current volume $\mathcal{V}_0$ should be very large, and $\varepsilon_{Q_1},~\varepsilon_{E_1}$ and $\varepsilon_{m_1}$ are of same order, one would expect that $\Omega_{Q_1}\ll \Omega_{E_1} \ll \Omega_{m_1}$, and in practice we can just ignore $\Omega_{Q_1}$ and $\Omega_{E_1}$ for small red shift $z$. When the expansion is modified (for example by including a second GFT mode), such that 
\begin{IEEEeqnarray}{rCl}
	H(H_0) &=& H_{\mathrm{single}}(H_0) + \delta H , \label{eq:deltaHdef}
\end{IEEEeqnarray} 
the preferred value of $H_0$ will change, and the deviation can be given through the response function $\mathcal{R}_{H_0}$ as we can see below.

\subsection{Deviation $\delta H$ in the presence of the second mode}
\begin{figure}[htpb]
	\centering
        \subfigure[~Approximated result]{\label{fig:deltaHpara38dataapp}\includegraphics[width=0.45\textwidth]{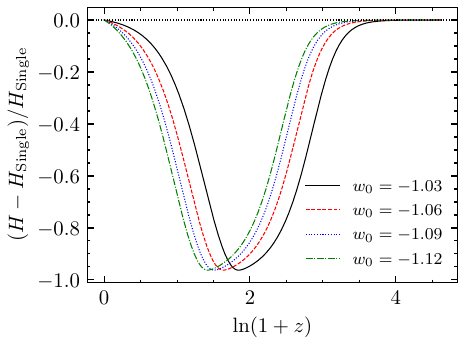}}
        \subfigure[~Exact result]{\label{fig:deltaHpara38datanum}\includegraphics[width=0.45\textwidth]{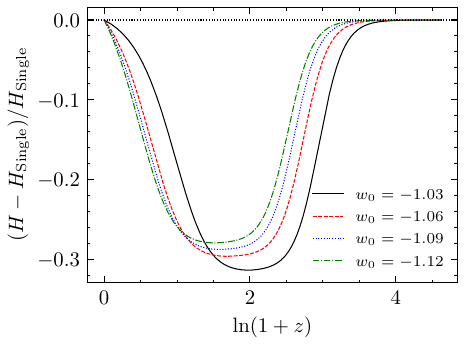}}
	\caption{The deviation $\delta H$ in the presence of the second mode. \ref{fig:deltaHpara38dataapp}: obtained from the approximated $w$ \eqref{eq:wvapprozw0}, corresponds to the black line of figure \ref{fig:wlnvpara38exaappcomp}; \ref{fig:deltaHpara38datanum}: obtained from the exact result of $w$ (calculated numerically), corresponds to red dashed line of figure \ref{fig:wlnvpara38exaappcomp}. The numerical value of $w$ improves the accuracy of the result, but the qualitative result remains the same. Data are the same as in figure \ref{fig:wlnvpara38exaappcomp}.}
	\label{fig:deltaHpara38data}
\end{figure}

In the presence of the second mode, the expansion history of our universe is modified. Effectively, such modification can be viewed as adding a phantom matter with \acrshort{eos} \eqref{eq:wvappro}~\cite{Oriti:2021rvm}, which can be rewritten using redshift $z$ as
\begin{IEEEeqnarray}{rCl}
	w &=& -1 - \frac{b}{\mathcal{V}_0}(1+z)^3	 ,\nonumber \\ 
&=& -1 + (1+w_0)(1+z)^3, \label{eq:wvapprozw0}
\end{IEEEeqnarray}
where $\mathcal{V}_0$ is the current volume and redshift $z$ satisfies $(1+z)^3=\mathcal{V}_0/\mathcal{V}$, and 
\begin{IEEEeqnarray}{rCl}
	w_0 &=& -1-\frac{b}{\mathcal{V}_0} , \label{eq:w0V0relation}
\end{IEEEeqnarray} 
is the current value of \acrshort{eos}. 
Then the modified expansion history in the presence of the second mode has the form 
\begin{IEEEeqnarray}{rCl}
	H^2&=& H_0^2\left[\Omega_{Q_1}(1+z)^{12}+\Omega_{E_1}(1+z)^9+\Omega_{m_1}(1+z)^6+\Omega_{d}\ee^{\int_0^z\frac{3(1+w(z'))}{1+z'}\dd z'}\right], \label{eq:Hubbletwomodes}
\end{IEEEeqnarray}
with \acrshort{eos} $w(z')$ has the form of \eqref{eq:wvapprozw0}.
Then, the relative modification $\displaystyle \delta H/H_{\mathrm{single}}$ is (for simplicity, we ignore terms proportional to $\Omega_{Q_1}$ and $\Omega_{{E_1}}$, since they are small)
\begin{IEEEeqnarray}{rCl}
	\frac{\delta H}{H_{\mathrm{single}}} &=& \frac{H-H_{\mathrm{single}}}{H_{\mathrm{single}}}, \nonumber \\
										 &=& -1 +\sqrt{\frac{\Omega_{m_1}(1+z)^6+\Omega_{\lambda_1}\exp\left\{(1+w_0)\left[(1+z)^3-1\right]\right\}}{\Omega_{m_1}(1+z)^6+\Omega_{\lambda_1}}} . \label{eq:deltaHHsingle}
\end{IEEEeqnarray} 
In the presence of the second mode, the universe will enter a phantom phase, thus we have $1+w_0<0$, which means $\delta H/H_{\mathrm{single}}$ is negative, as shown in figure \ref{fig:deltaHpara38dataapp}.
Furthermore, the deviation is non-vanishing only around its minimal value, determined by
\begin{IEEEeqnarray*}{rCl}
	\frac{\dd }{\dd z}\frac{\delta H}{H_{\mathrm{single}}} &\approx&  \frac{3(1+w_0)\Omega_{\lambda_1}(1+z)^2 \left\{\Omega_{m_1}(1+z)^3\left[(1+z)^3-2\right]-\Omega_{\lambda_1}\right\}}{2 \left[\Omega_{m_1}(1+z)^6+\Omega_{\lambda_1}\right]^2} =0 , 
\end{IEEEeqnarray*} 
where we used the fact that $1+w_0$ is small. In our model, $\Omega_{m_1}$ is usually much smaller than $\Omega_{\lambda_1}$\footnote{This is acceptable as one would expect that the contribution from free massless scalar field (which has \acrshort{eos} $w=1$) should vanish faster than the radiation (which has $w=1/3$), and the latter can already be ignored nowadays.}, hence the solution to the last equation can be approximated as
\begin{IEEEeqnarray}{rCl}
	z_{\mathrm{min},\delta H} &=& \left(\frac{\Omega_{\lambda_1}}{\Omega_{m_1}}\right)^{\frac{1}{6}}-1 . \label{eq:zmindeltaH}
\end{IEEEeqnarray} 
In the next subsection, we show that this value is close to the minimal position of the so-called response function $\mathcal{R}_{H_0}$ (see equation \eqref{eq:RH0def} for definition), hence the modification introduced by including a second mode will indeed change $H_0$.

Our approximation \eqref{eq:wvapprozw0} of $w$ deviates from the exact one quickly when $z$ is large (as can be seen from figure \ref{fig:wlnvpara38exaappcomp}), which results in a not that small deviation $\delta H$, as can be seen from figure \ref{fig:deltaHpara38dataapp}. From figure \ref{fig:deltaHpara38datanum}, we see that using the exact behaviour of $w$ instead, obtained by numerical methods, indeed improves the accuracy, and results in a smaller deviation. However, the qualitative feature remains the same, and the approximation \eqref{eq:wvapprozw0} is useful when we try to determine the minimal point of the deviation, which is given by equation \eqref{eq:zmindeltaH}. %

More generally, as we show in the appendix \ref{app:otherpara}, using other sets of parameters would not change the qualitative behaviour of the deviation $\delta H/H_{\mathrm{Single}}$ or the response function $\mathcal{R}_{H_0}$, only the quantitative details are modified.

Let us also emphasize that in our model the choice of $w_0$ is not independent of other parameters. In fact, according to the relation \eqref{eq:w0V0relation} between $w_0$ and $\mathcal{V}_0$, the value of $w_0$ will determine $\mathcal{V}_0$ as well, which is used in defining the redshift $z$. In other words, the choice of $w_0$ will determine what we call `now' in our model. Correspondingly, the value of relative energy density, such as $\Omega_{m_1}$ and $\Omega_{\lambda_1}$ will also change; this is due to the adoption of a modified expansion history, but to the change of different reference point for $z=0$. This is different from the theoretical context of ~\cite{Heisenberg:2022gqk}, where the definition of `now' is given once and for all, and correspondingly we only need one response function $\mathcal{R}_{H_0}$.

\subsection{The response function $\mathcal{R}_{H_0}$} \label{sec:responsefuctionRH0}
Now, for a small deviation of the expansion history $\delta H$ with the form of \eqref{eq:deltaHdef},
the preferred value of $H_0$ when fitting with observed data should be changed such that~\cite{Heisenberg:2022gqk}
\begin{IEEEeqnarray}{rCl}
	H(H_0+\delta H_0) &=& H_{\mathrm{single}}(H_0) +\Delta H . \label{eq:DeltaHdef}
\end{IEEEeqnarray} 
{Even though equation \eqref{eq:DeltaHdef} looks similar to \eqref{eq:deltaHdef}, they have different meanings. In fact, equation \eqref{eq:deltaHdef} refers to the modification of the expansion history compared to the single mode case $H_{\mathrm{single}}(H_0)$ (due to the inclusion of an additional mode in our GFT model, for example), indicated by the modification $\delta H$ which leaves the current value $H_0$ of the Hubble parameter unchanged. On the other hand, in equation \eqref{eq:DeltaHdef}, we take into account the fact that the value of $H_0$ will also be modified when we fit the modified expansion history to the observed data. After clarifying this point, we can now extract the $\delta H_0$ (the modification of $H_0$) from the modified expansion history.} At leading order, we have
\begin{IEEEeqnarray}{rCl}
	\frac{\Delta H}{H_{\mathrm{single}}} &=&\frac{H_0^2}{H_{\mathrm{single}}^2}\frac{\delta H_0}{H_0} + \frac{\delta H}{H_{\mathrm{single}}}. \label{eq:DeltaHvar}
\end{IEEEeqnarray} 
And in general, every cosmological quantity $g(z)$ will have the variation as following~\cite{Heisenberg:2022gqk}
\begin{IEEEeqnarray}{rCl}
	\frac{\Delta g(z)}{g(z)} &=&I_g(z)\frac{\delta H_0}{H_{\mathrm{single}}} +\int_0^{\infty}\frac{\dd x_z}{1+x_z}R_g(x_z,z)\frac{\delta H(x_z)}{H_{\mathrm{single}}(x_z)}  . \label{eq:Igzdef}
\end{IEEEeqnarray} 
In particular, for angular diameter distance~\cite{Heisenberg:2022gqk}
\begin{IEEEeqnarray}{rCl}
	d_A(z) &=& \frac{1}{1+z}\int_0^{z}\frac{\dd z}{H(z)}. 
\end{IEEEeqnarray} 
{To the first order of the variation, we have 
\begin{IEEEeqnarray}{rCl}
	\Delta d_A(z) &=& \frac{1}{1+z} \int_0^{z}\frac{\dd z}{H(H_0+\delta H_0)}-\frac{1}{1+z} \int_0^{z}\frac{\dd z}{H_{\mathrm{single}}(H_0)}, \nonumber\\ 
    &=&\frac{1}{1+z} \int_0^{z}\frac{\dd z}{H_{\mathrm{single}}(z)+\Delta H(z)}-\frac{1}{1+z} \int_0^{z}\frac{\dd z}{H_{\mathrm{single}}(z)}, \nonumber\\ 
    &=& -\frac{1}{1+z}\int_0^z\frac{\Delta H}{(H_{\mathrm{single}}+\Delta H)H_{\mathrm{single}}}\dd z, \nonumber\\ 
    &\approx& -\frac{1}{1+z}\int_0^z\frac{\Delta H}{H_{\mathrm{single}}}\frac{1}{H_{\mathrm{single}}}\dd z,\nonumber\\ 
    &=& -\frac{1}{1+z}\int_0^z\frac{H_0^2}{H_{\mathrm{single}}^3}\frac{\delta H_0}{H_0}\dd z  - \frac{1}{1+z}\int_0^z\frac{1}{H_{\mathrm{single}}}\frac{\delta H}{H_{\mathrm{single}}}\dd z, 
\end{IEEEeqnarray} 
where in the $4$-th line we have used the variation \eqref{eq:DeltaHvar}. Substituting the result into the equation \eqref{eq:Igzdef}, we obtain, }
\begin{IEEEeqnarray}{rCl}
	I_{d_A}(z) &=& - \frac{1}{\chi(z)} \int_0^z\dd x_z \frac{H_0^2}{H_{\mathrm{single}}^3}  ,\label{eq:IdAzdef} \\
	R_{d_A}(z)&=& -(1+x_z)\frac{\theta(z-x_z)}{\chi(z)H_{\mathrm{single}}(x_z)}, \label{eq:RdAzdef} 
\end{IEEEeqnarray} 
where $\chi(z)$ is the conformal distance
\begin{IEEEeqnarray}{rCl}
	\chi(z) &=& \int_0^z \frac{\dd x_z}{H_{\mathrm{single}}(x_z)} .  \label{eq:chizdef}
\end{IEEEeqnarray} 
Substituting equation \eqref{eq:chizdef} and the FLRW equation for single mode \eqref{eq:FLRWHsingle} into equations \eqref{eq:IdAzdef} and \eqref{eq:RdAzdef}, we see that both $I_{d_A}$ and $R_{d_A}$ are independent of $H_0$.

\begin{figure}[htpb]
	\centering
	\includegraphics[width=0.6\textwidth]{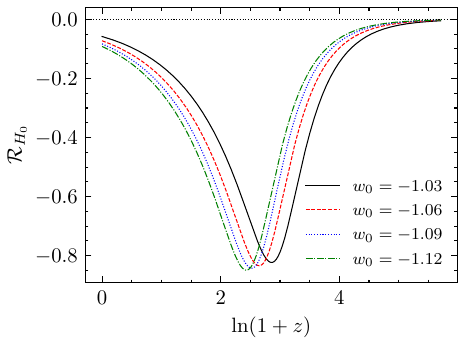}
	\caption{The response function $\mathcal{R}_{H_0}$. The data is the same as in figure \ref{fig:wlnvpara38exaappcomp}. Note that for $w_0=-1.03$ we have $\mathcal{V}_0=1.296\times 10^8$. Then the Hubble parameter in Planck units is $H_0=3.5\times 10^{-4}$, much larger than the value inferred from the Planck data, which in Planck units reads $H_0=1.18\times 10^{-61}$ (in SI units the value is $H_0=67.4~km~s^{-1}~Mpc^{-1}$~\cite{Aghanim:2018eyx}). The exact estimation of $H_0$ in our model from \acrshort{cmb} data requires including the non-relativistic matter into GFT, which is out of reach for the moment. And we leave this issue for future work.} \label{fig:Rhln1pzpara38comp}
\end{figure}

To see how the inclusion of a second mode changes the value of $H_0$, we need a {quantity} that is fixed for both one- and two-modes cases. {At the current stage, the massless scalar field is the only matter that couple to our GFT model, and a significant event would be the end of the bounce scenario, where the scalar field starts to dominate the cosmological evolution (with \acrshort{eos} $w=1$)~\cite{Oriti:2021rvm}. } For example, we can consider the angular diameter distance {when bounce ends} (with redshift $z_*$)\footnote{Currently, it's unclear how to determine the exact value of $z_*$ in our GFT model, but the exact value of $z_*$ is not important as long as it's large enough, {because the inclusion of the second mode will only modify the late time cosmology~\cite{Oriti:2021rvm}}. In the following, we simply take $z_*=1000$.}, whose deviation due to the modification of the expansion history is
\begin{IEEEeqnarray*}{rCl}
	\frac{\Delta d_{A}^*}{d_{A^*}} &=& I^*_{d_A}\frac{\delta H}{H_0} +\int_0^{\infty} \frac{\dd x_z}{1+x_z} R_{d_A}^{*}\frac{\delta H}{H_{\mathrm{single}}} , 
\end{IEEEeqnarray*} 
where we write $d_A(z_*)$ (and similarly for $I_{d_A}^*$ and $R_{d_A}^*$) for simplicity.  {The value of} $d_A^*$ {is fixed by the bouncing scenario and} should not change no matter how we modify the expansion history, i.e., we require $\Delta d_A^*\simeq 0$. This provides us the variation of $\delta H_0$ due to the modification of expansion history~\cite{Heisenberg:2022gqk}
\begin{IEEEeqnarray}{rCl}
	\frac{\delta H_0}{H_0} &=& \int_0^{\infty}\frac{\dd x_z}{1+x_z}\mathcal{R}_{H_0}\frac{\delta H}{H_{\mathrm{single}}},  \label{eq:deltaH0RH0}
\end{IEEEeqnarray} 
where the response function
\begin{IEEEeqnarray}{rCl}
	\mathcal{R}_{H_0} &=& -\frac{R_{d_A}^*}{I_{d_A}^*} , \label{eq:RH0def}
\end{IEEEeqnarray} 
which is also independent of $H_0$ as same as $I_{d_A}^*$ and $R_{d_A}^*$.
Since $H(z)\geq0$ in the whole history of our universe, we see that $\chi(z)>0$ and $I_{d_A}<0,~R_{d_A}<0$, which results in $\mathcal{R}_{H_0}<0$ as well, as shown in figure \ref{fig:Rhln1pzpara38comp}. On the other hand, the appearance of the second mode introduces phantom crossing, which results in a negative $\delta H$. Therefore, the preferred value $H_0$ will increase compared to the single mode case. 

Before we move on, let us take a closer look at the behaviour of response function $\mathcal{R}_{H_0}$. For $x_z<z^*$, we have 
\begin{IEEEeqnarray*}{rCl}
	\frac{\dd}{\dd x_z}R_{d_A}^* &=& -\frac{1}{\chi^*}\left[\frac{1}{H_{\mathrm{single}}(x_z)}-\frac{1+x_z}{H_{\mathrm{single}}(x_z)^2}\frac{\dd}{\dd x_z}H_{\mathrm{single}}(x_z)\right] . 
\end{IEEEeqnarray*} 
The minimal value is determined by $\frac{\dd}{\dd x_z}R_{d_A}^*=0$, which requires 
\begin{IEEEeqnarray*}{rCl}
\frac{1}{H_{\mathrm{single}}(x_z)}-\frac{1+x_z}{H_{\mathrm{single}}(x_z)^2}\frac{\dd}{\dd x_z}H_{\mathrm{single}}(x_z)&=& 0 .  
\end{IEEEeqnarray*} 
Substituting the FLRW equation for single mode \eqref{eq:FLRWHsingle}, and ignore $\Omega_{Q_1}$ and $\Omega_{E_1}$ since they are small, we get 
\begin{IEEEeqnarray}{rCl}
	6\Omega_{m_1}(1+z)^6 &=& \Omega_{m_1}(1+z)^6 +\Omega_{\lambda_1}. 
\end{IEEEeqnarray} 
Therefore, at the minimal value of $\mathcal{R}_{H_0}$, we have the redshift 
\begin{IEEEeqnarray}{rCl}
	z_{\mathrm{min},\mathcal{R}_{H_0}} &=& \left(\frac{\Omega_{\lambda_1}}{5\Omega_{m_1}}\right)^{\frac{1}{6}}-1 ,  \label{eq:zminRH0}
\end{IEEEeqnarray}
around which, we have $\Omega_{\lambda_1}\simeq5\Omega_{m_1}(1+z)^6$. Therefore, when the response function $\mathcal{R}_{H_0}$ deviates from $0$ significantly, we see that in the FLRW equation \eqref{eq:FLRWHsingle} the contribution from matter term has the same order as the cosmological constant term. This is just where the universe enters the phantom phase and the contribution from the second mode becomes noticeable. This can also be seen from the fact that by comparing equations \eqref{eq:zmindeltaH} and \eqref{eq:zminRH0}
\begin{IEEEeqnarray*}{rCl}
	z_{\mathrm{min},\mathcal{R}_{H_0}} = \frac{1}{5^{1/6}} z_{\mathrm{min},\delta H}+\frac{1}{5^{1/6}}-1. 
\end{IEEEeqnarray*} 
We see that $z_{\mathrm{min},\mathcal{R}_{H_0}}$ and $z_{\mathrm{min},\delta H}$ are close to each other, which means there would be an overlap for the regions where $H$ deviates from $H_{\mathrm{single}}$ and $\mathcal{R}_{H_0}$ deviates from $0$ respectively. Therefore, the preferred Hubble parameter $H_0$ changes indeed according to equation \eqref{eq:deltaH0RH0}.

Substituting equations \eqref{eq:RH0def} and \eqref{eq:deltaHHsingle} into equation \eqref{eq:deltaH0RH0} we can get the change of preferred $H_0$ in the presence of the second mode. Several results for different $w_0$ are shown in table \ref{tab:deltaH0comp}. We see that, when including a second mode into our GFT cosmology model, the value of $H_0$ inferred from data increases. {Qualitatively, a second mode will introduce a phantom phase where the energy density increases, hence elevate $H_0$ above the $\Lambda$CDM prediction under the CMB’s constraints on the Hubble parameter.} In other words, the effective phantom dynamics produced by quantum gravity interactions in our GFT condensate cosmology can alleviate the Hubble tension by increasing $H_0$ inferred from \acrshort{cmb} data~\cite{Heisenberg:2022gqk}.

\begin{table}[htpb]
	\centering
	\caption{The deviation of $H_0$ in the presence of second mode. The $4$th column lists results obtained using $w$ in the approximated form \eqref{eq:wvapprozw0}, while the $5$th column lists the result obtained using numerical $w$ from~\cite{Oriti:2021rvm}.}
	\label{tab:deltaH0comp}

	\begin{tabular}{c|cccc}
		$w_0$ & $\mathcal{V}_0$ & $\Omega_{m_1}$ & \begin{tabular}{@{}c@{}}$\delta H_0/H_0$ \\ (approx) \end{tabular} & \begin{tabular}{@{}c@{}}$\delta H_0/H_0$ \\ (numerical) \end{tabular}  \\
		\hline
		$-1.03$ & $1.30\times 10^8$ & $1.79\times 10^{-8}$ & $0.700$ & $0.269$ \\
		$-1.06$ & $6.92\times 10^7$ & $6.27\times 10^{-8}$ & $0.706$ & $0.271$ \\
		$-1.09$ & $4.74\times 10^7$ & $1.34\times 10^{-7}$ & $0.711$ & $0.260$ \\
		$-1.12$ & $3.61\times 10^7$ & $2.30\times 10^{-7}$ & $0.715$ & $0.253$ \\
	\end{tabular}
\end{table}

We emphasize again that in GFT cosmology the choice of $w_0$ (the current value of \acrshort{eos}) will also change the current volume $\mathcal{V}_0$ and hence the energy density $\Omega_{m_1}$ of matter. This fact is also reflected in table \ref{tab:deltaH0comp}. 

\section{Conclusions}
\label{sec:summary}

In this paper, we investigated several cosmological consequences of the presence of two dynamical quantum geometric modes in GFT condensate cosmology in mean field approximation, or, equivalently, in a hydrodynamics on minisuperspace framework. 
The effects can be encoded in the effective \acrshort{eos} $w$. Based on earlier results we know that the two-modes dynamics produces a late-time phantom phase in the evolution of the universe~\cite{Oriti:2021rvm}. 

The first result has been to determine the location of the phantom crossing, i.e. when $w$ crosses the line $w=-1$. Noting that phantom crossing must happen between the end of the Friedman phase and the minimum point of $w$, we showed that these two points of cosmic evolution are close to each other. This allows us to determine that, in our model, the universe will enter the de Sitter regime quickly after the phantom crossing, hence, if we are experiencing a phantom phase, the phantom crossing must happen recently with a low redshift. 

In the de Sitter regime, the evolution is dominated by a single mode, and our second result was to extract the precise value of the effective cosmological constant \eqref{eq:Lambdaqv} as a function of the microscopic parameters of our quantum gravity (\acrshort{gft}) model. Interestingly, from the analysis of the resulting expression, we see that a non-vanishing $\Lambda$  implies a bounce in the very early universe, replacing the cosmological singularity. Therefore, $\Lambda$ can be viewed as a remnant of the history of the universe in the far past.

As a third main result, using the method from~\cite{Heisenberg:2022gqk}, we have shown how the second mode changes the preferred value of $H_0$ when we fit data using our \acrshort{gft} cosmology model. We did it by approximately model the effects of the second quantum geometric mode in terms of an effective phantom matter, and showing that the deviation $\delta H$ is negative (figure \ref{fig:deltaHpara38data}) and that $\delta H/H$ is non-vanishing only around its minimum point, which is also close to the minimum point of response function $\mathcal{R}_{H_0}$, implying that the non-vanishing region of $\delta H/H$ and $\mathcal{R}_{H_0}$ overlap. This is important as otherwise the current value $H_0$ of the Hubble parameter will not change no matter how the expansion history \eqref{eq:Hubbletwomodes} is modified. Finally, we obtained the deviation $\delta H_0/H_0$ of the current Hubble parameter {$H_0$}, as shown in table \ref{tab:deltaH0comp}, which is positive as we expected. 

There are several ways to extend the scope of current work. We mention just a few. First, it is important to include in our model a more realistic matter content. Our universe is filled with non-relativistic matter, which has not been modelled yet in the \acrshort{gft} formalism. Without such inclusion, it is hard to compare the cosmological evolution predicted by the formalism with the actual cosmological data; once the inclusion is performed, on the other hand, the same procedure of current work can be followed for extracting actual cosmological predictions. Second, as we have seen that a non-vanishing cosmological constant implies a bounce in the far past of the evolution history; our analysis should be deepened to a more quantitative analysis of the precise details of the bouncing regime and its relation to the asymptotic de Sitter one.
{Third, we can consider more than two modes. Although, as mentioned earlier, the effects of additional modes on the phantom phase are expected to be small, it is still worthwhile to examine how modifications before the phantom phase may affect the preferred value of $H_0$.
}
Finally, we haven't taken into account the quantum corrections to the mean field dynamics we considered, and the renormalization of the underlying GFT quantum dynamics. On the one hand, {as} we have seen, the cosmological constant \eqref{eq:Lambdaqv} {depends} on the interaction coupling $\lambda_1$, which is subject to change due to the renormalization of the underlying quantum dynamics. A full renormalization analysis of quantum geometric models would provide more information on why $\Lambda$ is so small and about its evolution through cosmic history. Moreover, the asymptotic de Sitter regime lies in the dangerous regime of strong or at least non-negligible interactions, where the mean field hydrodynamic approximation is not obviously valid, and thus its robustness should be checked. {However, a full renormalization analysis is currently beyond reach for quantum geometric models, due to their analytic complexity. Thus, we have to resort to approximate methods, also in this respect. The results of \cite{Marchetti:2022nrf,Marchetti:2022igl} obtained via landau-Ginsburg methods offer both some reassurance in this respect and a basis for further developments: they support the picture of a condensate phase and a Gaussian critical point in the low-spin regime, and give evidence for the suppression of fluctuations also in the (weakly) interacting theory and the reliability of the mean field approximation also in this regime. More refined analyses, of course, remain necessary.}
Despite these limitations and necessary improvements, our results represent a concrete example of how quantum gravity can provide an explanation for large-scale cosmological puzzles, in an emergent spacetime scenario.

\acknowledgments
DO acknowledges support through the ATRAE Grant PR28/23 ATR2023-145735 funded by
MCIN /AEI /10.13039/501100011033, from the Deutsche Forschung Gemeinschaft via the Heisenberg Grant OR 432/3-1, from the John Templeton Foundation through the grant 62421, and from the Munich Center for Quantum Science and Technology (MCQST). XP was also supported by the China Scholarship Council and the Doctoral Initiation Grant 493150 from China West Normal University.

\appendix
\section{Numerical results from other sets of parameters}
\label{app:otherpara}

\begin{figure}[htpb]
	\centering
        \subfigure[Behaviour of $w$]{\label{fig:wlnvpara404142}\includegraphics[width=0.45\textwidth]{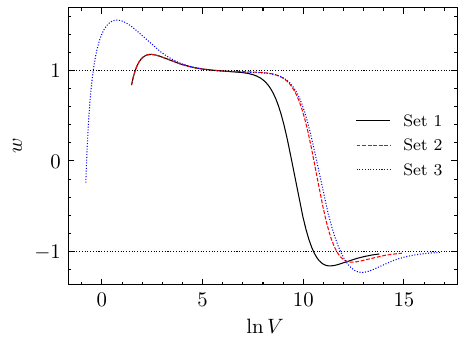}} \\
		\subfigure[Behaviour of $\delta H/H_{\mathrm{Single}}$]{\label{fig:deltaHpara404142datanum}\includegraphics[width=0.45\textwidth]{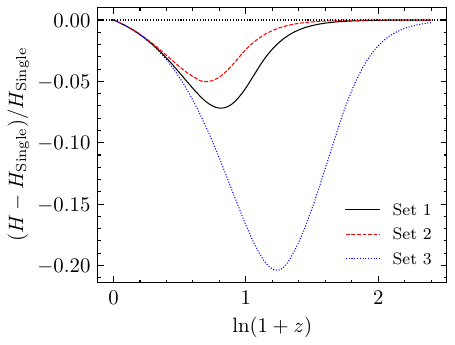}}
		\subfigure[Behaviour of $\mathcal{R}_{H_0}$]{\label{fig:Rhln1pzpara404142comp}\includegraphics[width=0.45\textwidth]{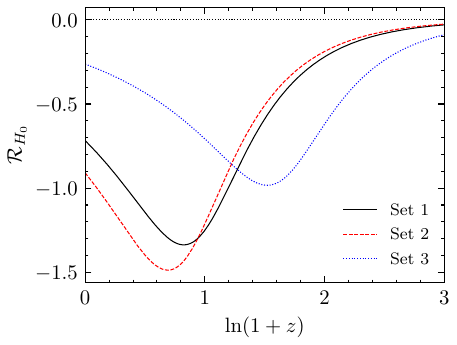}}
	\caption{The behaviour of the equation of state $w$, the deviation $\delta H/H_{\mathrm{Single}}$ and the response function $\mathcal{R}_{H_0}$ for different sets of parameters. In figure \ref{fig:deltaHpara404142datanum} and \ref{fig:Rhln1pzpara404142comp} we choose $w_0=-1.03$. Parameters in set $1$ are $\mathcal{V}_1=1/3,m_1^2=30,~E_1=50,~Q_1^2=900,\mathcal{V}_2=1/2,~m_2^2=20,~E_2=90,~Q_2^2=225,~\lambda_1=-10^{-7}, ~\lambda_2=-9.5\times10^{-7}$, such that $\phi_{1\infty}=0.9083$ and $\phi_{2\infty}=0.9663$. Parameters in set $2$ are $\mathcal{V}_1=1/3,m_1^2=30,~E_1=50,~Q_1^2=900,\mathcal{V}_2=1/2,~m_2^2=20,~E_2=90,~Q_2^2=225,~\lambda_1=-10^{-8}, ~\lambda_2=-9.5\times10^{-8}$, such that $\phi_{1\infty}=1.0134$ and $\phi_{2\infty}=1.0949$. While parameters in set $3$ are $\mathcal{V}_1=1/3,m_1^2=3,~E_1=5,~Q_1^2=9,\mathcal{V}_2=1/2,~m_2^2=2,~E_2=9,~Q_2^2=2.25,~\lambda_1=-10^{-9}, ~\lambda_2=-9.5\times10^{-9}$, such that $\phi_{1\infty}=3.4553$ and $\phi_{2\infty}=3.5311$.}
	\label{fig:wdeltaHRhpara404142}
\end{figure}

Figure \ref{fig:wdeltaHRhpara404142} shows the numerical results for other sets of parameters. For the deviation of the Hubble parameter and the response function, we fix the current value of the equation of state to be $w_0=-1.03$ but change the underlying \acrshort{gft} parameters.
We see that the qualitative behaviour of $w$, $\delta H/H_{\mathrm{Single}}$ and $\mathcal{R}_{H_0}$ are similar to each other, only the quantities details changed. Therefore, we can expect that our analysis in the main text is valid in general.

\nocite{*}

\providecommand{\href}[2]{#2}\begingroup\raggedright\endgroup

\end{document}